\documentclass[journal]{IEEEtran}

\ifCLASSINFOpdf
\else
\fi
\usepackage{dblfloatfix}    

\usepackage{cite}
\usepackage{amsmath,amssymb,amsfonts}
\usepackage{bbm}

\usepackage{graphicx}
\usepackage{epstopdf}
\usepackage{textcomp}
\usepackage{xcolor}
\usepackage{colortbl} 

\usepackage{url} 

\usepackage{bm}
\usepackage[normalem]{ulem} 
\usepackage{supertabular}
\usepackage{amsthm}
\theoremstyle{definition}

\newtheorem{remark}{Remark}
\usepackage[linesnumbered,ruled,vlined]{algorithm2e}
\usepackage{pbox}
\usepackage{float} 
\usepackage{subcaption} 

\usepackage{boldline,multirow}
\usepackage{array}

\usepackage{pbox}

\usepackage{dblfloatfix}

\newcommand*\diff{\mathop{}\!\mathrm{d}}

\newcounter{probNum}

\begin{document}
\title{Distributed Resource Allocation Optimization for User-Centric Cell-Free MIMO Networks}
%
%
\author{Hussein~A.~Ammar\IEEEauthorrefmark{1},~\IEEEmembership{Student Member,~IEEE}, 
	Raviraj~Adve\IEEEauthorrefmark{1},~\IEEEmembership{Fellow,~IEEE},
	Shahram~Shahbazpanahi\IEEEauthorrefmark{2}\IEEEauthorrefmark{1},~\IEEEmembership{Senior Member,~IEEE},
	Gary~Boudreau\IEEEauthorrefmark{3},~\IEEEmembership{Senior Member,~IEEE},
	and~Kothapalli~Venkata~Srinivas\IEEEauthorrefmark{3},~\IEEEmembership{Member,~IEEE}
	\thanks{This work was supported in part by the Natural Sciences and Engineering Research Council (NSERC) of Canada and in part by Ericsson Canada.}
	\thanks{
		\IEEEauthorrefmark{1}H. A. Ammar and R. Adve are with the Edward S. Rogers Sr. Department of Electrical and Computer Engineering, University of Toronto, Toronto, ON M5S 3G4, Canada (e-mail: ammarhus@ece.utoronto.ca; rsadve@comm.utoronto.ca).
	}
	\thanks{
		\IEEEauthorrefmark{2}S. Shahbazpanahi is with the Department of Electrical, Computer, and Software Engineering, University of Ontario Institute of Technology, Oshawa, ON L1H 7K4, Canada. He also holds a Status-Only position with the Edward S. Rogers Sr. Department of Electrical and Computer Engineering, University of Toronto.
	}
	\thanks{
		\IEEEauthorrefmark{3}G. Boudreau and K. V. Srinivas are with Ericsson Canada, Ottawa, ON K2K 2V6, Canada.
	}
}


%
%

\maketitle 

\makeatletter
\def\tagform@#1{\maketag@@@{\normalsize(#1)\@@italiccorr}}
\makeatother

\newcommand*{\CUGainComparedToDULongTerm}{$1.3$}
\newcommand*{\CUGainComparedToDUInstantanousRounded}{$1.8$}

\begin{abstract}
We develop two distributed downlink resource allocation algorithms for user-centric, cell-free, spatially-distributed, multiple-input multiple-output (MIMO) networks. In such networks, each user is served by a subset of nearby transmitters that we call distributed units or DUs. 
The operation of the DUs in a region is controlled by a central unit (CU). Our first scheme is implemented at the DUs, while the second is implemented at the CUs controlling these DUs. We define a hybrid quality of service metric that enables distributed optimization of system resources in a proportional fair manner. Specifically, each of our algorithms performs user scheduling, beamforming, and power control while accounting for channel estimation errors. Importantly, our algorithm does not require information exchange amongst DUs (CUs) for the DU-distributed (CU-distributed) system, while also smoothly converging. Our results show that our CU-distributed system provides \CUGainComparedToDULongTerm- to \CUGainComparedToDUInstantanousRounded-fold network throughput compared to the DU-distributed system, with minor increases in complexity and front-haul load - and substantial gains over benchmark schemes like local zero-forcing. We also analyze the trade-offs provided by the CU-distributed system, hence highlighting the significance of deploying multiple CUs in user-centric cell-free networks.
\end{abstract}

%
\IEEEpeerreviewmaketitle

\section{Introduction}
Scalability requires a system to accommodate growing demands gracefully~\cite{bondi2000Scalabilitycharacteristics}. Scalability is critical motivation for user-centric, cell-free, spatially-distributed, multiple-input multiple-output (MIMO) networks~\cite{Scalable9174860, interdonato2019ubiquitous} , where users can be served by many transmitters~\cite{cellFreeVersusSmallCells7827017, cellFreeUserCentricPower8901451}, denoted herein as distributed units (DUs). In these networks, a serving cluster of DUs is defined for each user based on a metric, e.g., channel power~\cite{cellFreeUserCentricPower8901451} or serving distance~\cite{PDPUsercentricVsDisjoint8969384}; each user is effectively located at the center of its serving cluster, thereby eliminating conventional cell edges. 

Designing flexible distributed resource allocation schemes for cell-free networks is still an open issue~\cite{cellFreeStochasticGeometry8379438} due to the difficulties arising from the lack of a regular cell structure and the overlapping serving clusters for the users. Theoretically, centralized resource allocation, wherein all transmissions are jointly optimized, provides the performance upper bound. However, such a scheme implies an enormous overhead in terms of real-time exchange of channel state information (CSI) and computation load. Furthermore, since the optimization problems involved are usually non-convex, even effective solutions may not lead to the global optimum.

A distributed system, on the other hand, provides a trade-off between performance and scalability. Compared to a centralized scheme, it provides an advantage due to lower front-haul load as well as lower computational and storage complexity per network node. Importantly, it allows for lower communication overhead with limited exchange of CSI. Hence, a distributed system is more practical to deploy than a centralized one~\cite{differentCooperationLevels8845768}. On the other hand, such an approach \emph{may} provide worse performance than a centralized scheme~\cite{8761828}, because the resource allocation is performed without a global view of the network and with limited coordination. 

A major challenge in designing a distributed resource allocation scheme is that the crucial signal-to-interference-plus-noise ratio (SINR) metric is coupled between all the transmitters. Hence, SINR-based approaches, like in~\cite{FR8310563, Ahmad9084256}, are not suitable for distributed resource allocation. One alternative is to use metrics like the signal-to-leakage-plus-noise ratio (SLNR)~\cite{SLNRprecoding4202177} which decouples the allocation problem between the different transmitters. However, SLNR does not allow for effective power allocation because the beam power scales equally in both the signal and leakage terms. Here, we modify the approach proposed for cellular networks~\cite{SLINR} based on a mix of  of inter-cell leakage and intra-cell interference. Specifically, for the cell-free MIMO network at hand, the combinatorial approach to user scheduling in~\cite{SLINR} is not possible. Additionally, unlike most works in distributed processing, we include fairness as a key criterion for user-scheduling, thereby avoiding repeatedly serving the same users with strongest channels.

The available efforts to implement a distributed resource allocation scheme in cell-free networks have been based on algorithms that require iterative exchange of signals between the base stations~\cite{9107496}. The investigation in~\cite{9238440} studies a distributed framework for cooperative precoding in cell-free MIMO requiring over-the-air signaling between the base stations instead of front-haul$\backslash$back-haul signaling. Similarly, the work in~\cite{8307115} maximizes the weighted sum rate (WSR) within joint transmission clusters without centralized beamformer processing. However, the problem is solved through an equivalent minimum square error-based problem that requires a feedback channel to update weights during algorithm iterations.

The authors in~\cite{PowerAlloc8630677} maximize the uplink minimum SINR by decoupling the problem into two sub-problems, one that designs the receiver filter and another that optimizes the power allocation, using an alternating optimization. The study in~\cite{LocaPartialZFBF9069486} propose two distributed variants of zero-forcing (ZF); however, this work does not consider user scheduling. The authors of~\cite{EE9212395} optimize the energy efficiency in the uplink of cell-free massive MIMO networks under different scenarios of signal quantization. The investigations in~\cite{Scalable9174860} and~\cite{bjornson2019scalable9064545} study a suboptimal but scalable power control policy that uses large-scale fading decoding. Moreover, the study in~\cite{dualDecomposition8901196} uses dual decomposition and the gradient method for resource allocation in a relaying system by relaxing the binary scheduling variables. Furthermore, the work in~\cite{differentCooperationLevels8845768} analyzes the uplink spectral efficiencies under four different levels of cooperation in cell-free implementations. However, this work does not optimize resource allocation, but rather numerically studies the performance under minimum mean-square error (MMSE) combining.

Distributed resource allocation can also be tackled through a game-theoretic framework. The studies in~\cite{DistribResourceAllo7676375, FullyDistResourceAllo6630117} use a distributed form of auction theory for user scheduling in conventional networks. Briefly, the users compete for the resources through bidding and assignment phases~\cite{bertsekas1979distributed}. One disadvantage of such approaches is the considerable communications required between the network nodes before an allocation occurs, hence leading to a substantial overhead. Furthermore, these have been developed for simple schemes like carrier sense multiple access.

The DUs that connect to users are, themselves, controlled by central units (CUs). While some models use a single CU to control all DUs~\cite{8761828, bjornson2019scalable9064545}, this would limit the scalability of the network. As in~\cite{LocaPartialZFBF9069486}, we consider a system of multiple central units (CUs), each controlling the DUs in its region. Under the user-centric cell-free MIMO scheme at hand, a serving cluster of DUs is defined specifically for each user. In contrast to the schemes available in the literature, we develop two \emph{totally} distributed resource allocation schemes that perform user scheduling, beamforming, and implicit power control in a user-centric MIMO network. The first solution is implemented on the DUs, while the second one is deployed on the CUs\footnote{Although this study is not concerned with the core-network protocols that should be used to deploy a distributed CU system under the user-centric cell-free network architecture, the authors note that distributed software defined network (SDN)~\cite{distributedSDN8187644} is a promising avenue to implement the flexible network architecture suggested in this paper.}. We acquire \emph{local} CSI between each DU and its users, an approach shown to be scalable~\cite{channelAcquisition5462883}. We then define a weighted pseudo-rate function that depends on a hybrid leakage and intra-(DU$\backslash$CU) interference; the weights implement user fairness. Our objective decouples the problem between the different network nodes, with each node making decisions independent of the other nodes in the network.

To solve our formulated problem, we employ tools such as block coordinate descent, fractional programming, and compressive sensing to develop an algorithm that converges smoothly in a non-decreasing manner. In addition to standard CSI estimation, we further propose using either statistical CSI or the spatial traffic distribution to compute the leakage needed to construct our objective function. These alternative approaches further reduce any required information exchange amongst processing nodes. Notably, our results show that the proposed methods to compute the leakage are very effective with a dense distribution of users. 

To the best of the authors' knowledge, this work is one of the first studies to consider fully distributed resource allocation, and is the first study which investigates fully distributed user-scheduling and beamforming (not only power control) for user-centric cell-free MIMO networks. In most of the literature, user scheduling is neglected and the users are assumed pre-selected. Hence, our work fills two gaps by focusing on distributed schemes and on user scheduling. Our scheme is different from the literature, e.g.,~\cite{9238440}, in the sense that it conducts resource allocation without using feedback channels between the network entities. Additionally, to maintain scalability, we use a network with distributed CUs~\cite{8761828, bjornson2019scalable9064545}.

A few works in the literature focus on scalability. The study in~\cite{8761828} proposes a simple sub-optimal power allocation using conjugate beamforming (to be able to perform a distributed beamforming), but this study does not optimize user scheduling or beamforming. In~\cite{Scalable9174860} the authors study the uplink of a single-CU network, using large-scale fading decoding and power control; however, the scheme is not distributed. The authors of~\cite{d2021user} employ a combinatorial search for user-association based on the position of the access points. The system uses conjugate beamforming and maximizes the sum rate, however, this scheme is also not distributed. The scheme sacrifices performance to enable a low fronthaul load. Thus, our developed schemes are fundamentally different from those found in literature.

The theory developed here sets the foundation for deployment of an optimized resource allocation scheme. Specifically, the contributions of our paper are:
\begin{itemize}
	\item Developing two distributed resource allocation schemes that optimize resources, including user scheduling and beamforming (not only power allocation) in a cell-free, user-centric, MIMO network. These schemes are based on a hybrid leakage and interference metric that eliminates the need for information exchange and implements fairness amongst users.
	\item Proposing and testing three approaches to calculate the leakage term in the metric. These approaches require different levels of computational complexity and required real-time information while producing comparable performance, especially at high user densities.
	\item Analyzing the computation complexity of our two proposed approaches, illustrating the substantial reduction in complexity compared to a centralized resource allocation scheme, with, in some cases, improved performance.
	\item Highlighting the importance of a multiple-CU user-centric cell-free network by illustrating trade-offs in performance and front-haul load 
	provided by the CU-distributed system compared to DU-distributed and centralized systems.
\end{itemize}

The rest of the paper is organized as follows. Section~\ref{section:Model} presents the system model. Section~\ref{sec:DUsystem} formulates the resource allocation problem in a DU-distributed system and develops the steps required to solve this problem. Section~\ref{sec:CUsystem} casts the problem as a CU-distributed system and develops an approach to perform the resource allocation. Section~\ref{section:scalability_leakage} proposes further methods to enhance the scalability of calculating the leakage term. Section~\ref{section:results} reports on our numerical results and findings. Finally, Section~\ref{section:conclusion} concludes our discussion.

\emph{Notation:} a lower or upper case letter, e.g., $a$ or $A$, represents a scalar, a bold lower case letter, e.g., ${\bf a}$, represents a vector, while a bold upper case letter, e.g., ${\bf A}$, represents a matrix. The term 
$[{\bf A}]_{ij}$ is the $(i,j)$th entry of matrix ${\bf A}$. The operators $(\cdot)^{-1}$, $(\cdot)^T$ and $(\cdot)^H$, used as superscripts, denote the inverse, transpose, and conjugate transpose, respectively. $\|\cdot\|_2$ and $|\cdot|$ are the vector and scalar Euclidean norms ($\ell_2$ norm), $\|\cdot\|_p$ is the $\ell_p$ norm, and ${\bf I}_m$ is the $m$-dimensional identity matrix. Calligraphic letters, e.g., $\mathcal{A}$, are used to indicate sets with its cardinality represented by $|\mathcal{A}|$. We use ${\bf A} = [\mathcal{A}]$ to construct a matrix or vector using the elements of set $\mathcal{A}$. The spaces $\mathbb{B}$, $\mathbb{C}^m$, and $\mathbb{C}^{m\times n}$ represent the set of binary numbers, complex $m\times 1$ column vectors, and complex $m\times n$ matrices, respectively. $\mathbb{E}\{\cdot\}$ is the expectation operator, and $\mathbf{x} \sim \mathcal{CN}(\mathbf{m},\mathbf{R})$ indicates that $\mathbf{x}$ is a complex Gaussian random vector with mean $\mathbf{m}$ and covariance $\mathbf{R}$.

\section{System Model}\label{section:Model}
\subsection{Network and Signal Model}
\begin{figure}
	\centering
	\includegraphics[width=1\linewidth]{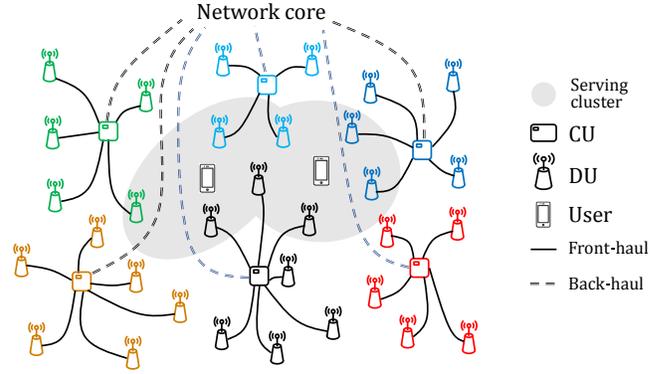}
	\vspace{-0.5em}
	\caption{User-centric cell-free network with \emph{either} DU-distributed or CU-distributed processing.}
	\label{fig:UserCentricClustering}
	\vspace{-1em}
\end{figure}

We consider a user-centric cell-free MIMO network, which operates in time-division duplex (TDD) mode. Our network, illustrated in Fig.~\ref{fig:UserCentricClustering}, comprises $Q$ CUs, each controlling $N$ DUs represented by the set $\mathcal{B}_q$. We denote the region containing the DUs in $\mathcal{B}_q$ as a \emph{virtual cell}. Accordingly, we have a total of $NQ$ DUs represented by the set of sets $\mathcal{B}=\{\mathcal{B}_1,\ \mathcal{B}_2, \dots, \mathcal{B}_Q\}$. Each DU is equipped with $M$ antennas, while each user is equipped with a single antenna. We use $\mathcal{U}$ to represent the set of users that need to be served, where $|\mathcal{U}|$ is a random number, but is much larger than the number of available resources. The DUs serve users coherently\footnote{Coherent transmission requires phase synchronization across the DUs, which can be achieved through synchronization protocols like the IEEE 1588v2 (Precision Time Protocol PTP)~\cite{9120376}. The system can achieve this synchronization by properly choosing the serving clusters and the use of a cyclic prefix~\cite{PDPUsercentricVsDisjoint8969384}. We note that synchronization issues are out of the scope of this paper; they need to be studied in a dedicated work using different tools from those used in this paper.}. 

Based on this model, for each user $u$, we define a serving cluster $\mathcal{C}_u$  comprising the DUs that \emph{can potentially} serve the user. Specifically, here we define $\mathcal{C}_u$ based on the criterion $\{ \psi_{ru}\beta(d_{ru}) \ge \rho : r \in \mathcal{B}\}$, where the term $\psi_{ru}$ accounts for the (lognormal) shadowing, $\beta(d_{ru})$ accounts for the path loss, which depends on the distance $d_{ru}$ between DU $r$ and user $u$. If no DU can meet this connection criterion, $\mathcal{C}_u$ comprises the DU providing the highest average received power, i.e., largest $\psi_{ru}\beta(d_{ru})$. Based on this scheme, we can formally define $\mathcal{C}_u = \{r: \psi_{ru}\beta(d_{ru}) \ge \rho\} \cup \{{\rm arg}\ \max_r\ \psi_{ru}\beta(d_{ru})\}$. The set $\mathcal{C}_u$, therefore excludes DUs that cannot contribute significantly to the user's useful signal while also limiting the number of users each DU serves. We assume a low-mobility profile for the users and hence we use the popular block fading model for the channel. Hence, the channel (small-scale and large-scale fading) is assumed static within each channel coherence interval. The large-scale fading also changes much more slowly than the small-scale fading, and hence stays constant for many coherence intervals. The block fading model is widely accepted and used in the literature when high-mobility scenarios are not being studied~\cite{cellFreeVersusSmallCells7827017} as the case of this paper.

We develop two different distributed algorithms for resource allocation. The first variant runs on the DUs, while the second variant runs on the CUs. The developed algorithms provide user scheduling and perform resource allocation. These procedures include channel estimation, formation of serving clusters, user scheduling, beamforming, and fairness control for the users. Finally, it is worth noting that we will be using the \emph{global} indices $q$, $r$, and $u$ to, respectively, refer to the CUs, DUs and users found in the network.

In the downlink, the signal received by user $u \in \mathcal{U}$ can be written as
\begin{align}\label{eq:signalModel}
y_{u} =&
\displaystyle
\sum_{r \in \mathcal{C}_u} \sqrt{{s}_{ru}}
{\bf h}_{ru}^H {\bf w}_{ru} x_{u}
\nonumber \\
&
+ \sum_{u' \in \mathcal{U}, u' \ne u} \sum_{r' \in \mathcal{C}_{u'}} \sqrt{{s}_{r'u'}} {\bf h}_{r'u}^H {\bf w}_{r'u'} x_{u'}
+ z_{u}.
\end{align}
The first term in \eqref{eq:signalModel} is the useful signal, the second term is the interference, and the third is the additive white Gaussian noise (AWGN) $z_{u}\sim\mathcal{CN}(0, \sigma_z^2)$. The scalar ${s}_{ru} \in \mathbb{B}$ is the scheduling decision at DU $r$ for user $u$ (that is, when $s_{ru} = 1$, DU $r$ schedules user $u$, otherwise $s_{ru} = 0$), ${\bf h}_{ru}\in \mathbb{C}^{M \times 1}$ is the channel vector between the two peers, ${\bf w}_{ru}$ is the beamforming weight or the transmit precoder used by DU $r$ to serve the user with an overall power budget of $p$ for the DU
, and $x_{u}$ is the zero-mean data symbol intended for user $u$ with $\mathbb{E}\{|x_{u}|^2\} = 1$.

We model the channel between DU $r$ and user $u$ as ${\bf h}_{ru} = \sqrt{\psi_{ru} \beta(d_{ru})} {\bf g}_{ru} \in \mathbb{C}^{M \times 1}$, where the term ${\bf g}_{ru} \sim\mathcal{CN}({\bf 0},{\bf I}_M)$ accounts for the small-scale fading, and as noted earlier, $\psi_{ru}$ and $\beta(d_{ru})$ account for lognormal shadowing and path loss, respectively. We define the set $\mathcal{E}_{r}$ representing the users that need to be served by DU $r$. These sets $\{\mathcal{E}_{r}: r \in \mathcal{B}\}$ can obtained directly from the sets $\{\mathcal{C}_u : u \in \mathcal{U} \}$. Crucially, because of user-centric clustering, each $\mathcal{E}_{r}$ can partially or totally overlap with other $\mathcal{E}_{r'}$, for $r' \ne r$; similarly, the same user may ``belong'' to multiple CUs.

\subsection{Channel Estimation}
For a TDD system, channel estimation is performed in the uplink and the estimated channel is used in downlink assuming channel reciprocity. In the pilot-training phase of length $\tau_p$, the signal ${\bf Y}_r \in \mathbb{C}^{M \times \tau_p}$ observed at DU $r$ can be written~as
\begin{align}\label{eq:receivedPilotSig}
{\bf Y}_r = \sum_{u \in \mathcal{U}} \sqrt{p_u} {\bf h}_{ru} {\bm \phi}_u + {\bf Z}_r,
\end{align}
where ${\bm \phi}_u \in \mathbb{C}^{1 \times \tau_p}$ is the unit norm (${\bm \phi}_u {\bm \phi}_u^H = 1$) pilot sequence used by user $u$, $p_u$ is the transmit power, and ${\bf Z}_r$ is the noise with entries distributed as $\mathcal{CN}\left(0, \sigma_Z^2 \right)$. Following the assumptions in~\cite{cellFreeUserCentricPower8901451, cellFreeVersusSmallCells7827017}, we assume the knowledge of the the large-scale fading and the transmit power used.

Unfortunately, using pilots that are orthogonal among the users, i.e., ${\bm \phi}_u {\bm \phi}_{u'}^H = 0, \forall u' \ne u$, requires having $\tau_p \ge |\mathcal{U}|$ which is unlikely to be feasible. We assume that the users are grouped such that the pilot sequences inside each group are orthogonal, while the same pilot sequences are used across groups. 
Specifically, as in~\cite{ammarC_RA_UC}, we use the hierarchical agglomerative clustering (HAC) algorithm to cluster the users into subsets each containing a number of users less than or equal to $\tau_p$, the length of the pilot sequence. The available orthogonal pilot sequences are randomly assigned to the users inside each subset. The intuition is to keep users sharing the same pilot sequence as far as possible from each other, thus minimizing pilot contamination in our user-centric cell-free network~\cite{randomVsStructuredPilots8403508}. Additionally, the choice of HAC is based on its consistency and its lack of sensitivity to the choice of the distance-metric used to construct the subsets~\cite{karypis2000comparison}.

We can extract the channel of user $u$ at DU $r$ by first defining ${\bf \breve{y}}_{ru} = \frac{1}{\sqrt{p_u}} {\bf Y}_r {\bm \phi}_u^H$, hence eliminating all users' contributions other than the ones using the same pilot sequence ${\bm \phi}_u$. The channel $\{{\bf h}_{ru} : u \in \mathcal{U}$\} can then be estimated using linear MMSE as~\cite{kay1993fundamentals}
\begin{align}
{\bf \hat{h}}_{ru} = 
{\bf D}_{ru} \left( \sum_{u' \in \mathcal{U}_u} {\bf D}_{ru'} + \frac{\sigma_Z^2}{p_u} {\bf I}_M \right)^{-1} {\bf \breve{y}}_{ru},
\end{align}
where ${\bf D}_{ru} \in \mathbb{C}^{M \times M}$ is a diagonal matrix with entries $\left[{\bf D}_{ru}\right]_{mm} \triangleq \psi_{ru} \beta(d_{ru})$, and $\mathcal{U}_u$ are the users using the same pilot sequence as user $u$ (including user $u$).

\begin{remark}
	Due to path loss, each DU $r$ only needs to estimate the channel vectors of nearby users. 
	In Section~\ref{section:scalability_leakage}, we show that each DU need only estimate the channels to users $u \in \mathcal{E}_r$, i.e., its own users and it can use large-scale fading statistics or a traffic distribution model for the other users, thus enhancing the scalability of the required channel estimation.
\end{remark}

The estimated channel ${\bf \hat{h}}_{ru}$ is distributed according to $\mathcal{CN}\left( {\bf 0}, {\bm \Psi}_{ru} \right)$, with ${\bm \Psi}_{ru}$ defined as
\begin{align}
{\bm \Psi}_{ru} \triangleq {\bf D}_{ru} \left( \sum_{u' \in \mathcal{U}_u} {\bf D}_{ru'} + \frac{\sigma_Z^2}{p_u} {\bf I}_M \right)^{-1} {\bf D}_{ru} .
\end{align}
It is known from the theory of MMSE estimation that the channel estimation error $\mathrm{ {\bf e}}_{ru} = {\bf h}_{ru} - {\bf \hat{h}}_{ru}$ is 
distributed as $\mathcal{CN}\left({\bf 0}, {\bm \Theta_{ru}}\right)$, where ${\bm \Theta_{ru}} \triangleq {\bf D}_{ru} - {\bm \Psi}_{ru}$~\cite{kay1993fundamentals}.

Based on this, the model for the signal received at user $u$ can be written as
\begin{align}\label{eq:signalModel_imperfectChan}
y_{u} &=
\sum_{r \in \mathcal{C}_u} \sqrt{{s}_{ru}}
{\bf \hat{h}}_{ru}^H {\bf w}_{ru} x_{u}
+ \sum_{r \in \mathcal{C}_u} \sqrt{{s}_{ru}} \mathrm{ {\bf e}}_{ru}^H {\bf w}_{ru} x_{u}
\nonumber \\
& \ \ 
+ \sum_{u' \in \mathcal{U}, u' \ne u} \sum_{r' \in \mathcal{C}_{u'}}
\sqrt{{s}_{r'u'}} {\bf \hat{h}}_{r'u}^H {\bf w}_{r'u'} x_{u'}
\nonumber \\
& \ \ 
+ \sum_{u' \in \mathcal{U}, u' \ne u} \sum_{r' \in \mathcal{C}_{u'}}
\sqrt{{s}_{r'u'}} \mathrm{ {\bf e}}_{r'u}^H {\bf w}_{r'u'} x_{u'}
+ z_{u} .
\end{align}
Including the random estimation error, $\mathrm{ {\bf e}}_{ru}$, in~\eqref{eq:signalModel_imperfectChan} allows us to implement robust beamforming that exploits the error covariance matrix ${\bm \Theta_{ru}}$ and hence compensate for some of the error.

We now define hybrid expressions that account for both the leakage and intra-(DU$\backslash$CU) interference. For distributed implementation, these expressions use \textit{locally constructed} beamformers.

\section{DU-Distributed System}\label{sec:DUsystem}
\subsection{Hybrid Leakage-Interference}
Let $\mathcal{U}_{-r} = \mathcal{U} \backslash \mathcal{E}_r$, represent the set of users that do not belong to $\mathcal{E}_r$, i.e., $\mathcal{U}_{-r}  = \{u \mid u \notin \mathcal{E}_r \}$. We define an expression for the average power (leakage) experienced by the users $\mathcal{U}_{-r}$ from DU $r$ by serving user $u$. Averaged over the random channel estimation error this power is given by
\begin{align}\label{eq:leakage_DUsystem}
	&L_{ru}\left({\bf w}_{ru}\right)
	\nonumber \\
	&=
	\mathbb{E}_{{\bf e}}\Bigg\{
	\sum_{u' \in \mathcal{U}_{-r}} t_{r,u'} \left| {\bf \hat{h}}_{ru'}^H  {\bf w}_{ru} \right|^2
	+ \sum_{u' \in \mathcal{U}_{-r}} t_{r,u'} \left| \mathrm{ {\bf e}}_{ru'}^H {\bf w}_{ru} \right|^2
	\Bigg\}
	\nonumber
	\\
	&\  
	=
	\sum_{u' \in \mathcal{U}_{-r}} t_{r,u'} \left| {\bf \hat{h}}_{ru'}^H  {\bf w}_{ru} \right|^2
	+ \sum_{u' \in \mathcal{U}_{-r}} t_{r,u'} {\bf w}_{ru}^H {\bm \Theta}_{ru'} {\bf w}_{ru} .
\end{align}
The vector ${\bf \hat{h}}_{ru'}$ for $u' \in \mathcal{U}_{-r}$ is the estimated \emph{leakage} channel between DU $r$ and user $u' \in \mathcal{U}_{-r}$. The term $t_{r,u'} \in \mathbb{B}$ is defined at DU $r$ and represents the \textit{assumption} about user $u' \in \mathcal{U}_{-r}$ being scheduled by at least one of its serving DUs $r' \in \mathcal{C}_{u'}$.

We also define our optimization variables as the matrix ${\bf W}_r = \left[ \left\{{\bf w}_{ru} : u \in \mathcal{E}_r \right\} \right] \in \mathbb{C}^{M \times |\mathcal{E}_r|}$, and ${\bf s}_r = \left[ \left\{ s_{ru} : u \in \mathcal{E}_r \right\} \right]^T \in \mathbb{B}^{|\mathcal{E}_r| \times 1}$ the beamformers and scheduling variables used by DU $r$. For each pair of DU $r$ and user $u$, we now define a hybrid signal-to-leakage-and-intra-DU-interference-and-noise ratio (SLINR-D) as
\begin{align}\label{eq:SLNR_R}
\xi_{ru}
\triangleq
\frac{ s_{ru}
	{\bf w}_{ru}^H {\bf \hat{h}}_{ru} {\bf \hat{h}}_{ru}^H {\bf w}_{ru} 
}
{ 
	A_{ru}\left({\bf s}_r, {\bf W}_{r} \right)
},
\end{align}
\begin{figure*}[b]
	\hrule
	\begin{align}\label{eq:SLINR_subterm}
		&A_{ru}\left({\bf s}_r, {\bf W}_{r} \right)
		=
		\resizebox{0.82\textwidth}{!}
		{$ \displaystyle
			L_{ru}\left({\bf w}_{ru}\right)
			+
			\mathbb{E}_{{\bf e}}\Bigg\{
			s_{ru} \left| \mathrm{ {\bf e}}_{ru}^H {\bf w}_{ru} \right|^2
			+ \sum_{u' \in \mathcal{E}_r, u' \ne u} s_{ru'} \left| {\bf \hat{h}}_{ru}^H {\bf w}_{ru'} \right|^2
			+ \sum_{u' \in \mathcal{E}_r, u' \ne u} s_{ru'} \left| \mathrm{ {\bf e}}_{ru}^H {\bf w}_{ru'} \right|^2
			+ \sigma_z^2
			\Bigg\}
			$}
		\nonumber \\
		& \ 
		=
		\resizebox{0.88\textwidth}{!}
		{$ \displaystyle
			L_{ru}\left({\bf w}_{ru}\right)
			+
			s_{ru} {\bf w}_{ru}^H {\bm \Theta}_{ru} {\bf w}_{ru}
			+ \sum_{u' \in \mathcal{E}_r, u' \ne u} s_{ru'} {\bf w}_{ru'}^H {\bf \hat{h}}_{ru} {\bf \hat{h}}_{ru}^H {\bf w}_{ru'}
			+ \sum_{u' \in \mathcal{E}_r, u' \ne u} s_{ru'} {\bf w}_{ru'}^H {\bm \Theta}_{ru} {\bf w}_{ru'}
			+ \sigma_z^2
			$}
		.
	\end{align}
\end{figure*}
where $A_{ru}\left({\bf s}_r, {\bf W}_{r} \right)$, used to simplify the notation, is the leakage plus intra-DU interference and noise averaged over the unknown random channel estimation error, and it is defined in~\eqref{eq:SLINR_subterm}.

In~\eqref{eq:SLNR_R} we treat channel estimation error $\mathrm{ {\bf e}}_{ru}$ as additional noise with covariance ${\bm \Theta_{ru}}$ as defined earlier. The first term in~\eqref{eq:SLINR_subterm} is the power leakage to the users not served by DU $r$, the second is the self-interference resulting from imperfect CSI of the serving channel, the third and fourth terms are the intra-DU interference experienced by user $u$, and the final term is the noise power.

\begin{remark}
	It is worth commenting on the role of self-interference in~\eqref{eq:SLINR_subterm}. Previous works based on leakage, do not mix in interference. However, from~\eqref{eq:SLNR_R} and~\eqref{eq:SLINR_subterm}, using purely leakage means that the beam power ($||\mathbf{w}_{ru}||^2$) would effectively cancel out in all but the noise term. This effectively takes away the role of power allocations, to the detriment of performance~\cite{SLINR}.
\end{remark}
\begin{remark}
Even for a user served by a \textit{cluster of DUs}, we can still define an SLINR-D for each DU-user pair. Also,~\eqref{eq:SLNR_R} is still indirectly coupled across the scheduling decisions of the other DUs through the terms $\{t_{r,u'} : u' \in \mathcal{U}_{-r}\}$. In centralized resource allocation, the scheduling of users is performed by a single entity, hence we can simply set $t_{r,u} = \min\left(1, \sum_{r' \in \mathcal{C}_{u}} s_{r'u} \right)$. However, in a DU-distributed system, a DU cannot know which users are scheduled by the other DUs, complicating the development of a distributed scheduling algorithm. In Section~\ref{section:scalability_leakage}, we address this issue by introducing different methods to calculate the leakage.
\end{remark}

Based on~\eqref{eq:SLNR_R}, we define a weighted \emph{pseudo-rate} between each DU $r$ and its user $u \in \mathcal{E}_r$ as
\begin{align}\label{eq:WPR}
{\rm WPS}_{ru}
\triangleq
\delta_{u}
\log\left( 1 + \xi_{ru}
\right).
\end{align}
As the name suggests, the weighted pseudo-rate employs the SLINR-D metric in a rate-like function while providing for fairness. We will optimize this pseudo-rate. Here, $\delta_{u}$ is a term that accounts for the proportional fairness of user $u$. At time slot $i+1$, these weights are the inverse of the long-term exponentially averaged achieved data rate, i.e., $\delta_{u}^{(i+1)} = \frac{1}{\bar{R}_u^{(i)}}$~\cite{yu2011adaptive}, where $\bar{R}_u^{(i)}  = \eta R_u^{(i)} + (1 - \eta) \bar{R}_u^{(i-1)}$ is the user's exponentially weighted rate averaged over the previous time slots. Here, $\ 0 < \eta < 1$ acts as a forgetting factor, and $R_u^{(i)}$ is the data rate at time slot $i$.

The motivation to optimize the weighted pseudo-rate is, in fact, two-fold. First, by only using local variables, it allows for the development of a DU-distributed system. Second, it acknowledges that maximizing the useful signal while minimizing the leakage and intra-DU interference is a prudent practice. Although the SLINR lacks an operational meaning (from an engineering viewpoint) still it can be interpreted as a proxy to enhance the performance~\cite{SLNRprecoding4202177}. Specifically, when a DU $r$ minimizes the leaked interference to users $\mathcal{U}_{-r}$, it minimizes the interference terms experienced by these users, and hence it enhances the SINR for these users. The same applies for the intra-DU interference experienced by users $\mathcal{E}_r$. The mix between the leakage and the intra-DU interference instead of using the leakage only prevents the power of the beam of the DU from scaling equally in both the signal and leakage terms when it is being optimized. In our numerical results, we compare our approach to an SINR-based approach which, unfortunately, requires centralized deployment, and we show that our approach is effective.

\subsection{Problem Definition}
For each DU $r$, we define the following optimization problem:
\begin{subequations}\label{eq:Prob_eachDU_est}
	\begin{align}
	\stepcounter{probNum}
	(\mathrm{P\arabic{probNum}})(r) \quad
	\max_{ {\bf W}_r, {\bf s}_r }\quad & \sum_{u \in \mathcal{E}_r} \delta_{u}
	\log\left( 1 + 
	\xi_{ru}
	\right) & 
	\label{eq:Prob_eachDU_est_obj}
	\\
	%
	\text{s.t.}\quad 
	& \sum_{u \in \mathcal{E}_r} s_{ru} \le M
	&
	\label{eq:Prob_eachDU_est_cap}
	\\
	&
	\sum_{u\in \mathcal{E}_r} ||{\bf w}_{ru}||_2^2 \le p
	\label{eq:Prob_eachDU_est_power}
	&
	\\
	&
	\xi_{ru}
	=
	\frac{ s_{ru}
		{\bf w}_{ru}^H {\bf \hat{h}}_{ru} {\bf \hat{h}}_{ru}^H {\bf w}_{ru} 
	}
	{ 
		A_{ru}\left({\bf s}_r, {\bf W}_{r} \right)
	},
	&
	u \in \mathcal{E}_r
	\label{eq:Prob_eachDU_est_form}
	\\
	&
	s_{ru} \in \{0, 1\},
	&
	u \in \mathcal{E}_r
	\label{eq:Prob_eachDU_est_binary}
	%
	.
	\end{align}
\end{subequations}
The term $A_{ru}\left({\bf s}_r, {\bf W}_{r} \right)$ is defined in~\eqref{eq:SLINR_subterm}. We note that, in \eqref{eq:Prob_eachDU_est_form}, $\{ t_{r,u'} : u' \in \mathcal{U}_{-r} \}$ are not decision variables, but rather they represent the \textit{assumptions} of DU $r$ about the scheduled users served by other DUs. When this knowledge is not available by any means, $\{ t_{r,u'} : u' \in \mathcal{U}_{-r} \}$ are simply set to $1$. In Section~\ref{section:scalability_leakage}, we evaluate alternative methods to handle this important issue.

The aim in~\eqref{eq:Prob_eachDU_est} is to optimize, at DU $r$, the decision variables ${\bf s}_r$ and ${\bf W}_r$, respectively the user scheduling and beamforming vectors. The constraint in~\eqref{eq:Prob_eachDU_est_cap} restricts the number of users served by DU $r$ to the number of antennas $M$,~\eqref{eq:Prob_eachDU_est_power} specifies the power budget of DU $r$, \eqref{eq:Prob_eachDU_est_form} treats the SLINR-D expression as an auxiliary variable, and \eqref{eq:Prob_eachDU_est_binary} shows that a user may be either scheduled or not. It is well established that problems having the form of~\eqref{eq:Prob_eachDU_est} are NP-hard~\cite{Complexity4453890}, thus obtaining the global optimum is computationally prohibitive, and only a local optimum can be obtained. The problem is  mixed-integer and non-convex due to the binary variables and their presence in both the numerator and denominator of the utility function.


Using fractional programming, the problem in~\eqref{eq:Prob_eachDU_est} can be reformulated and written as
\vspace{-0.5em}
\begin{subequations}\label{eq:Prob_eachDU_est_reform}
	\begin{align}
	\stepcounter{probNum}
	(\mathrm{P\arabic{probNum}})(r) \quad
	\max_{ {\bf W}_r, {\bf s}_r, {\bm \xi}_r, {\bm \zeta}_r }\quad & 
	f_2\left( {\bf s}_r, {\bf W}_r, {\bm \xi}_r, {\bm \zeta}_r \right)
	\label{eq:Prob_eachDU_est_reform_obj}
	& 
	\\
	\text{s.t.}\quad 
	& \sum_{u \in \mathcal{E}_r} s_{ru} \le M
	&
	\\
	&
	\sum_{u\in \mathcal{E}_r} ||{\bf w}_{ru}||_2^2 \le p
	\label{eq:Prob_eachDU_est_reform_power}
	&
	\\
	&
	s_{ru} \in \{0, 1\},
	&
	u \in \mathcal{E}_r .
	%
	\end{align}
\end{subequations}
with the function $f_2\left( {\bf s}_r, {\bf W}_r, {\bm \xi}_r, {\bm \zeta}_r \right)$ found in~\eqref{eq:Prob_eachDU_est_reform_obj} is defined~as
\begin{align}\label{eq:Linearized_R}
	&f_2\left( {\bf s}_r, {\bf W}_r, {\bm \xi}_r, {\bm \zeta}_r \right) =
	\sum_{u \in \mathcal{E}_r}
	\delta_{u} \left( \log\left( 1 + \xi_{ru} \right) - \xi_{ru} \right)
	\nonumber \\
	&\ \ 
	+ \sum_{u \in \mathcal{E}_r}
	\bigg(
	2 \text{Re}\left\{
	\zeta_{ru}^{*}
	\sqrt{\delta_{u}\left( 1 + \xi_{ru}\right)}
	s_{ru}
	{\bf w}_{ru}^H {\bf \hat{h}}_{ru}
	\right\}
	\nonumber \\
	&\ \ 
	-
	|\zeta_{ru}|^2
	\left(
	s_{ru}
	{\bf w}_{ru}^H {\bf \hat{h}}_{ru} {\bf \hat{h}}_{ru}^H {\bf w}_{ru}
	+ A_{ru}\left({\bf s}_r, {\bf W}_{r} \right)
	\right)
	\bigg)
	.
\end{align}
where ${\bm \zeta}_r \in \mathbb{C}^{|\mathcal{E}_r| \times 1}$ is a new auxiliary variable vector introduced by fractional programming~\cite{FR8310563}. Fractional programming refers to optimizing functions composed of ratios. The function can be composed of a single ratio, or, as in our case, a sum of ratios. The numerator and denominator can be nonlinear. Techniques used to obtain local optimum for such optimization problems include the Charnes-Cooper method~\cite{schaible1974parameter}, Dinkelbach's method~\cite{dinkelbach1967nonlinear}, and quadratic transform~\cite{FR8310563}.

The reformulation in~\eqref{eq:Prob_eachDU_est_reform} is based on the Lagrangian formulation and fractional programming. Please refer to Appendix~\ref{eq:DUsystem_Form1} for details.

In what follows, we obtain optimal expressions for the optimization variables, one set of variables at a time, i.e., we use block coordinate descent. As for the scheduling variables ${\bf s}_r$, we optimize them using a combinatorial search as described~below. 

When the variables other than $\zeta_{ru}$ are fixed, the first optimality condition of~\eqref{eq:Linearized_R} with respect to $\zeta_{ru}$ results in the optimal value~as
\begin{align}\label{eq:zeta_DU_est}
\zeta_{ru}
=
\frac{
	s_{ru}
	\sqrt{\delta_{u}\left( 1 + \xi_{ru}\right)}
	{\bf w}_{ru}^H {\bf \hat{h}}_{ru}
}
{
	s_{ru}
	{\bf w}_{ru}^H {\bf \hat{h}}_{ru} {\bf \hat{h}}_{ru}^H {\bf w}_{ru}
	+ A_{ru}\left({\bf s}_r, {\bf W}_{r} \right)
} .
\end{align}
We write the Lagrangian formulation of~\eqref{eq:Prob_eachDU_est_reform}, and when variables $({\bf s}_r, {\bm \xi}_r, {\bm \zeta}_r)$ are fixed, we derive the optimality condition to obtain the optimal expression for the transmit precoder ${\bf w}_{ru}$ as shown in~\eqref{eq:beamformer_DU_est}.
\begin{figure*}[b]
\hrule
\begin{align}\label{eq:beamformer_DU_est}
&{\bf w}_{ru}
=
s_{ru}
\zeta_{ru}^{*}
\sqrt{\delta_{u}\left( 1 + \xi_{ru}\right)}
\Bigg(
|\zeta_{ru}|^2
\Big(
s_{ru}
\left ({\bf \hat{h}}_{ru} {\bf \hat{h}}_{ru}^H + {\bm \Theta}_{ru}\right)
+ 
\sum_{u' \in \mathcal{U}_{-r}} t_{r,u'} {\bf \hat{h}}_{ru'} {\bf \hat{h}}_{ru'}^H
+ \sum_{u' \in \mathcal{U}_{-r} } t_{r,u'} {\bm \Theta}_{ru'}
\Big)
\nonumber \\
& \quad \quad \quad \quad 
+
s_{ru} \sum_{u' \in \mathcal{E}_r, u' \ne u} |\zeta_{ru'}|^2 \left( {\bf \hat{h}}_{ru'} {\bf \hat{h}}_{ru'}^H + {\bm \Theta}_{ru'} \right)
+ \mu_r {\bf I}_M
\Bigg)^{-1}
{\bf \hat{h}}_{ru}
\nonumber \\
&
=
\resizebox{0.92\textwidth}{!}
{$ \displaystyle
s_{ru}
\zeta_{ru}^{*}
\sqrt{\delta_{u}\left( 1 + \xi_{ru}\right)}
\Bigg(
|\zeta_{ru}|^2
\Big( 
\sum_{u' \in \mathcal{U}_{-r}} t_{r,u'} {\bf \hat{h}}_{ru'} {\bf \hat{h}}_{ru'}^H
+ \sum_{u' \in \mathcal{U}_{-r} } t_{r,u'} {\bm \Theta}_{ru'}
\Big)
+
s_{ru} \sum_{u' \in \mathcal{E}_r} |\zeta_{ru'}|^2 \left( {\bf \hat{h}}_{ru'} {\bf \hat{h}}_{ru'}^H + {\bm \Theta}_{ru'} \right)
+ \mu_r {\bf I}_M
\Bigg)^{-1}
{\bf \hat{h}}_{ru} ,
$}
\end{align}

\begin{subequations}\label{eq:scheduling_search}
	\begin{align}
		\stepcounter{probNum}
		(\mathrm{P\arabic{probNum}})(r) \quad
		\max_{ \{v_{ru,m}: u\in\mathcal{E}_r, 1 \leq m \leq M\} }\quad & 
		\sum_{m = 1}^{M}
		v_{ru,m}
		\sum_{u \in \mathcal{E}_r}
		\delta_{u}
		\log\left( 1 + 
		\frac{
			{\bf \widetilde{w}}_{rm}^H {\bf \hat{h}}_{ru} {\bf \hat{h}}_{ru}^H {\bf \widetilde{w}}_{rm}
		}
		{
			\widetilde{A}_{ru,m}
		}
		\right)
		\label{eq:HungarianObjective}
		\\
		\text{s.t.}\quad
		& \sum_{m = 1}^{M} v_{ru,m} \le 1,
		\quad \quad \quad \quad \quad \quad \quad \quad \quad \quad
		\quad \quad \quad \quad\ 
		u \in \mathcal{E}_r
		\label{eq:TotalProblem_scheduling_c_sum1_CP}
		\\
		& \sum_{u \in \mathcal{E}_r} v_{ru,m} = 1,
		\quad \quad \quad \quad \quad \quad \quad \quad \quad \quad
		\quad
		m = 1, \dots, M
		\label{eq:TotalProblem_scheduling_c_sum2_CP}
		\\
		&
		v_{ru,m} \in \{0, 1\},
		\quad \quad \quad \quad \quad
		\quad \quad \quad
		u \in \mathcal{E}_r, m = 1, \dots, M
		\label{eq:TotalProblem_scheduling_c_Binary_CP}
	\end{align}
\end{subequations}

\begin{align}\label{eq:ATilde}
	\widetilde{A}_{ru,m}
	=
	\sum_{u' \in \mathcal{U}, u' \notin \mathcal{E}_r} t_{r,u'} {\bf \widetilde{w}}_{rm}^H \left( {\bf \hat{h}}_{ru'} {\bf \hat{h}}_{ru'}^H + {\bm \Theta}_{ru'} \right) {\bf \widetilde{w}}_{rm}
	+ {\bf \widetilde{w}}_{rm}^H {\bm \Theta}_{ru} {\bf \widetilde{w}}_{rm}
	+ \sum_{m' = 1, m' \ne m}^{M} {\bf \widetilde{w}}_{rm'}^H \left( {\bf \hat{h}}_{ru} {\bf \hat{h}}_{ru}^H + {\bm \Theta}_{ru} \right) {\bf \widetilde{w}}_{rm'}
	+ \sigma_z^2 .
\end{align}
\end{figure*}

The variable $\mu_r$ denotes the Lagrange multiplier for the power budget constraint~\eqref{eq:Prob_eachDU_est_reform_power}, and is inversely related to the beamformers' power at DU $r$; $\mu_r$ can be determined through the complementary slackness condition for the power budget. Specifically, let ${\bf w}_{ru}(\mu_r)$ denote the right-hand side of~\eqref{eq:beamformer_DU_est}. If $\sum_{u\in \mathcal{E}_r} \| {\bf w}_{ru}(0)\|_2^2 \le p$, then $\mu_r = 0$, otherwise the value of $\mu_r$ can be determined through a bisection search to satisfy $\sum_{u\in \mathcal{E}_r} \| {\bf w}_{ru}(\mu_r)\|_2^2 = p$.

When the variables $({\bf W}_r, {\bm \xi}_r, {\bm \zeta}_r)$ are fixed, the problem of optimizing ${\bf s}_r$ at each DU $r$ is a combinatorial problem, where the users are matched to the available non-zero beams $\left\{ {\bf \widetilde{w}}_{rm}: m \in \left\{1,\ \dots,\ M \right\} \right\}$ found at the DU. Thus, the problem of optimizing the scheduling is cast as a combinatorial problem as follows.

Constraint~\eqref{eq:TotalProblem_scheduling_c_sum1_CP} states that a user can be assigned at most one non-zero beam, and constraint~\eqref{eq:TotalProblem_scheduling_c_sum2_CP} states that all non-zero beams are assigned. In~\eqref{eq:HungarianObjective}, $\widetilde{A}_{ru,m}$ is the term that contains the leakage, interference and noise terms resulting from assigning the non-zero beam $m$ on DU $r$ to user $u$, and is defined as~\eqref{eq:ATilde}.

It is important to note that, $\widetilde{A}_{ru,m}$ is not affected if the non-zero beams other than $m$ are assigned to different users, because \textit{in the scheduling phase} the beamformers are fixed.

\newcommand{\varPermMatrix}{This problem in~\eqref{eq:scheduling_search} is an agent-task assignment problem~\cite{bokhari2012assignment}, and can be written as finding an $(|\mathcal{E}_r| \times M)$ column-permutating matrix ${\bf V}_r$ that maximizes $({[1 \dots 1]{\bf A}_r{\bf V}_r^T [1 \dots 1]^T})$ for a given matrix ${\bf A}_r$. A column-permuting matrix is binary with a single 1 in each column in a unique location. Here, $[{\bf V}_r]_{um} = v_{ru,m}$ and the entries of ${\bf A}_r \in \mathbb{R}^{|\mathcal{E}_r| \times M}$ are the pseudo-rates in~\eqref{eq:HungarianObjective}  representing the utility of the users served by DU $r$. The matrix ${\bf V}_r$, hence, serves to assign each non-zero beam (column of ${\bf V}_r$) to a specific user (row of ${\bf V}_r$).}  
\varPermMatrix\ 

\begin{algorithm}[t]
	\SetAlgoLined
	\SetInd{0.1em}{1em}
	\caption{Distributed resource allocation at each DU $r$}
	\label{algorithm:SLNR_DU}
	Initialize ${\bf W}_r$ for all users. \label{step:initializeBeam}\\ 
	Initialize ${\bf s}_r$ by selecting $M$ users with max~\eqref{eq:WPR}.\\
	\While{ \textbf{NOT} converged}{\label{step:Algo_terminate_DU}
		Update $\{ \xi_{ru} : u \in \mathcal{E}_r \}$ using~\eqref{eq:Prob_eachDU_est_form}.
		\\
		Update $\{ \zeta_{ru} : u \in \mathcal{E}_r \}$ using~\eqref{eq:zeta_DU_est}.\\
		Update $\{ {\bf w}_{ru} : u \in \mathcal{E}_r\}$ using~\eqref{eq:beamformer_DU_est} and $\mu_r$ through the complementary slackness condition of power budget.
		\\
		Update ${\bf V}_r$ by solving~\eqref{eq:scheduling_search} using Hungarian algorithm, then $\resizebox{0.26\textwidth}{!}
		{$
			\{s_{ru} = \sum_{m = 1}^{M} v_{ru,m} : u \in \mathcal{E}_r\}
		$}$
		\label{step:determined_scheduling}
	}
\end{algorithm}

This assignment problem can be solved efficiently in polynomial time using the Hungarian algorithm~\cite{kuhn1955hungarian} (also known as the Kuhn–Munkres algorithm). \label{page:Hunagarian}\newcommand{\varHunagarian}{This algorithm executes a series of iterative manipulations for the rows and columns of the matrix ${\bf A}_r$ (or alternatively a cost matrix), which allows us to find the maximum (or minumum for cost matrix) entries for the assignment of the rows of ${\bf A}_r$ to the columns (agent-task assignment). Implementations for the Hungarian algorithm can be readily found in many scripting languages.}\varHunagarian

Finally, user $u$ is scheduled by DU $r$ if it is assigned a non-zero beam, i.e.,
$s_{ru} = \sum_{m = 1}^{M} v_{ru,m}$. After defining the optimal expression for each variable type when the other variables are fixed, we can construct Algorithm~\ref{algorithm:SLNR_DU} that uses block coordinate descent to optimize the resource allocation in a distributed fashion on each DU. The initialization of the beamforming in Step~\ref{step:initializeBeam} can be performed using conjugate beamforming, while Steps~\ref{step:Algo_terminate_DU}-\ref{step:determined_scheduling} optimize one variable at a time until convergence. We will analyze convergence in our section on numerical results.

\section{CU-Distributed System}\label{sec:CUsystem}
The previous section designed an algorithm that allows each DU to perform its own allocation decisions. However, if a CU can coordinate the DUs under its control, we can include the intra-CU interference, i.e., inside the virtual cell of each CU. A CU-distributed algorithm would, therefore, provide a balance between some coordination (and information exchange) and the completely centralized case where all CUs are jointly optimized. Additionally, the CU would decide on user scheduling for all the DUs under its control. We emphasize, however, that a user may be associated with DUs under the control of different CUs (user-centric clustering). A user may, therefore, be scheduled by one CU, but not by the other. For each CU $q$, we define the set of users $\bar{\mathcal{U}}_{q} \triangleq \underset{r' \in \mathcal{B}_q}{\cup} \mathcal{E}_{r'}$ that includes all the users connected to at least one DU under the control of CU $q$. Due to user-centric clustering, the different sets $\{\bar{\mathcal{U}}_{q}\}_{q=1}^{Q}$ overlap. 

For each user $u \in \bar{\mathcal{U}}_{q}$ and DUs $\mathcal{D}_{qu} = \left( \mathcal{C}_u \cap \mathcal{B}_q \right)$, we define the concatenation of beamformers, estimated channels, channel estimation error and scheduling variables as
{\allowdisplaybreaks
\begin{align}
{\bf \bar{w}}_{qu} &= \left[ \left\{ {\bf w}_{ru}^T : r \in \mathcal{D}_{qu} \right\} \right]^T
\in 
\mathbb{C}^{M|\mathcal{D}_{qu}| \times 1}
\label{eq:beamformer_concat_def}
\\
{\bf \bar{h}}_{qu,u'} &= \left[ \left\{ {\bf \hat{h}}_{ru'}^T : r \in \mathcal{D}_{qu} \right\} \right]^T
\in 
\mathbb{C}^{M|\mathcal{D}_{qu}| \times 1}
\label{eq:channel_concat_def}
\\
\mathrm{ {\bf \bar{e}}}_{qu,u'} &= \left[ \left\{ \mathrm{ {\bf e}}_{ru'}^T : r \in \mathcal{D}_{qu} \right\} \right]^T
\in 
\mathbb{C}^{M|\mathcal{D}_{qu}| \times 1}
\label{eq:channelError_concat_def}
\\
{\bf S}_{qu} &= \left({\rm diag} \left( \left\{ s_{ru} : r \in \mathcal{D}_{qu} \right\} \right) \otimes {\bf I}_M \right)
\in 
\mathbb{B}^{M|\mathcal{D}_{qu}| \times M|\mathcal{D}_{qu}|}
\label{eq:schduling_concat_def}
\end{align}
}
We note that~\eqref{eq:channel_concat_def} and~\eqref{eq:channelError_concat_def} represent the estimated channel and estimation error, respectively, between user $u'$ and the serving DUs $\mathcal{D}_{qu}$ for user $u$. As in the previous section, we define an expression for the power leakage experienced by the users in $\bar{\mathcal{U}}_{-q} = \mathcal{U} \backslash \bar{\mathcal{U}}_{q}$ from all the DUs due to serving user $u$. Averaged over the random channel estimation error this leakage is given by~\eqref{eq:leakage_CUsystem}, where the term $\mathbb{E}\left\{ \mathrm{ {\bf \bar{e}}}_{qu,u'} \mathrm{ {\bf \bar{e}}}_{qu,u'}^H \right\} = {\bf \bar{\bm \Theta}}_{qu,u'}$ in~\eqref{eq:leakage_CUsystem} represents the covariance of the channels' estimation error obtained from treating the unknown random Gaussian error as additional noise.
\begin{figure*}[b]
\hrule
\begin{align}\label{eq:leakage_CUsystem}
\bar{L}_{qu}\left( {\bf \bar{w}}_{qu} \right) &=
\mathbb{E}_{{\bf e}}\Bigg\{
\sum_{u' \in \bar{\mathcal{U}}_{-q}} \bar{t}_{q,u'} \left| {\bf \bar{h}}_{qu,u'}^H {\bf \bar{w}}_{qu} \right|^2
+ \sum_{u' \in \bar{\mathcal{U}}_{-q}} \bar{t}_{q,u'} \left| \mathrm{ {\bf \bar{e}}}_{qu,u'}^H {\bf \bar{w}}_{qu} \right|^2
\Bigg\}
\nonumber\\
& =
\sum_{u' \in \bar{\mathcal{U}}_{-q}} \bar{t}_{q,u'} \left| {\bf \bar{h}}_{qu,u'}^H {\bf \bar{w}}_{qu} \right|^2
+ \sum_{u' \in \bar{\mathcal{U}}_{-q}} \bar{t}_{q,u'} {\bf \bar{w}}_{qu}^H {\bf \bar{\bm \Theta}}_{qu,u'} {\bf \bar{w}}_{qu},
\end{align}
\end{figure*}

As before, we define the hybrid signal-to-leakage-and-intra-CU-interference-and-noise-ratio (SLINR-C) between each CU $q$ and user $u \in\bar{\mathcal{U}}_{q}$~as
\begin{align}
\bar{\xi}_{qu} & \triangleq
\frac{
	{\bf \bar{w}}_{qu}^H {\bf S}_{qu}^{1/2} {\bf \bar{h}}_{qu,u} {\bf \bar{h}}_{qu,u}^H {\bf S}_{qu}^{1/2} {\bf \bar{w}}_{qu} 
}
{ 
	B_{qu}\left(\mathcal{S}_q, \mathcal{W}_{q} \right)
},
\end{align}
where the leakage plus intra-CU interference and noise, averaged over the random estimation error, is defined as~\eqref{eq:SLINR_subterm_CU_v2}.
\begin{figure*}[b]
\begin{align}\label{eq:SLINR_subterm_CU_v2}
B_{qu}\left(\mathcal{S}_q, \mathcal{W}_{q} \right)
&=
\bar{L}_{qu}\left( {\bf \bar{w}}_{qu} \right)
+
\mathbb{E}_{{\bf e}}\Big\{
\left| \mathrm{ {\bf \bar{e}}}_{qu,u}^H {\bf S}_{qu}^{1/2} {\bf \bar{w}}_{qu} \right|^2
+ \sum_{u' \in \bar{\mathcal{U}}_{q}, u' \ne u} {\bf \bar{w}}_{qu'}^H {\bf S}_{qu'}^{1/2} 
{\bf \bar{h}}_{qu',u} {\bf \bar{h}}_{qu',u}^H {\bf S}_{qu'}^{1/2} {\bf \bar{w}}_{qu'}
\nonumber \\
& \quad
+ \sum_{u' \in \bar{\mathcal{U}}_{q}, u' \ne u} {\bf \bar{w}}_{qu'}^H {\bf S}_{qu'}^{1/2} \mathrm{ {\bf \bar{e}}}_{qu',u} \mathrm{ {\bf \bar{e}}}_{qu',u}^H {\bf S}_{qu'}^{1/2} {\bf \bar{w}}_{qu'}
+ \sigma_z^2
\Big\},
\nonumber \\
& =
\bar{L}_{qu}\left( {\bf \bar{w}}_{qu} \right)
+
{\bf \bar{w}}_{qu}^H {\bf S}_{qu}^{1/2} {\bf \bar{\bm \Theta}}_{qu,u} {\bf S}_{qu}^{1/2} {\bf \bar{w}}_{qu}
+ \sum_{u' \in \bar{\mathcal{U}}_{q}, u' \ne u} {\bf \bar{w}}_{qu'}^H {\bf S}_{qu'}^{1/2} 
{\bf \bar{h}}_{qu',u} {\bf \bar{h}}_{qu',u}^H {\bf S}_{qu'}^{1/2} {\bf \bar{w}}_{qu'}
\nonumber \\
& \quad
+ \sum_{u' \in \bar{\mathcal{U}}_{q}, u' \ne u} {\bf \bar{w}}_{qu'}^H {\bf S}_{qu'}^{1/2} {\bf \bar{\bm \Theta}}_{qu',u} {\bf S}_{qu'}^{1/2} {\bf \bar{w}}_{qu'}
+ \sigma_z^2
\end{align}
\end{figure*}

Here in~\eqref{eq:SLINR_subterm_CU_v2}, ${\bf S}_{qu}$ is defined in~\eqref{eq:schduling_concat_def} with diagonal entries $s_{ru} \in \mathbb{B}$ denoting the scheduling variable of user $u$ at DU $r$, and $\bar{t}_{q,u} \in \mathbb{B}$ is defined at CU $q$ to represent the \textit{assumption} for user $u$ being scheduled by at least one of its serving DUs $r' \notin \mathcal{B}_q$ not under the control of CU $q$. 

Similar to the previous section, for each CU $q$ we can optimize the pseudo-rates as~\eqref{eq:optProbCU}.
\begin{figure*}[b]
	\hrule
\begin{subequations}\label{eq:optProbCU}
	\begin{align}
	\stepcounter{probNum}
	(\mathrm{P\arabic{probNum}})(q) \quad
	\max_{ \mathcal{W}_{q}, \mathcal{S}_q }\quad & \sum_{u \in \bar{\mathcal{U}}_{q} } \delta_{u}
	\log\left( 1 + 
	\bar{\xi}_{qu}
	\right) 
	\\
	\text{s.t.}\quad 
	& \sum_{u \in \mathcal{E}_r} s_{ru} \le M,
	\quad
	r \in \mathcal{B}_q
	\\
	&
	\sum_{u\in \mathcal{E}_r} ||{\bf w}_{ru}||_2^2 \le p,
	\quad
	r \in \mathcal{B}_q
	\label{eq:TotalUtilityProblem_U_1_w_eachCU}
	\\
	&
	\bar{\xi}_{qu} =
	\frac{ 
		{\bf \bar{w}}_{qu}^H {\bf S}_{qu}^{1/2} {\bf \bar{h}}_{qu,u} {\bf \bar{h}}_{qu,u}^H {\bf S}_{qu}^{1/2} {\bf \bar{w}}_{qu} 
	}
	{ 
		B_{qu}\left(\mathcal{S}_q, \mathcal{W}_{q} \right)
	},
	\quad
	r \in \mathcal{B}_q, u \in \mathcal{E}_r
	\label{eq:SLINR_eachCU}
	\\
	&
	s_{ru} \in \{0, 1\},
	\quad
	r \in \mathcal{B}_q, u \in \mathcal{E}_r
	\end{align}
\end{subequations}
\end{figure*}
The set $\mathcal{W}_{q} = \{{\bf W}_{r} : r \in \mathcal{B}_q\}$ in~\eqref{eq:optProbCU} represents the matrix of beamformers used by the DUs $\mathcal{B}_q$, where, as noted earlier, ${\bf W}_r = \left[ \left\{{\bf w}_{ru} : u \in \mathcal{E}_r \right\} \right] \in \mathbb{C}^{M \times |\mathcal{E}_r|}$ are the beamformers used by each DU $r$ to serve its users. Similarly, the scheduling variables $\mathcal{S}_q = \{{\bf S}_{qu} : u \in \bar{\mathcal{U}}_{q} \}$, where ${\bf S}_{qu}$ contains the scheduling variables $s_{ru}$ by each DU $r \in \mathcal{D}_{qu}$. $\mathcal{W}_{q}$ and  $\mathcal{S}_q$ are the optimization variables.

Because of the coupling across DUs under the control of CU $q$, we require a new approach to the scheduling problem. The scheduling variables can be related to the indicator function of the beamformer which, in turn, can be written as an $\ell_0$-norm, i.e., $s_{ru} = \mathbbm{1} \{ \|{\bf w}_{ru}\|_2^2 \} = \| \| {\bf w}_{ru}\|_2^2 \|_0$. From compressive sensing~\cite{CompressedSensing1614066}, we can approximate the $\ell_0$-norm as a weighted $\ell_1$-norm that is easier to work with as $\|{\bf x}\|_0 \simeq \sum_{m} \alpha_m |x_m| = \| {\bm \alpha}^T {\bf x}\|_1,$~\cite{candes2008enhancing}. The variables $\alpha_{m}$ are weights that can be updated iteratively. For our problem, we can define the weights $\alpha_{ru}$~as
\begin{align}\label{eq:weightsUpdate}
\alpha_{ru} = \frac{1}{ \left\| {\bf w}_{ru} \right\|_2^2 + \epsilon} \ ,
\end{align}
where $\epsilon > 0$ provides stability and ensures that a zero-valued component in $\left\| {\bf w}_{ru} \right\|_2^2$ does not strictly prohibit a nonzero estimate in the next iteration; importantly, the results are not very sensitive to $\epsilon$ and it can be chosen to be slightly smaller than the expected value of $\left\| {\bf w}_{ru} \right\|_2^2$ for the scheduled users~\cite{candes2008enhancing}. 

Based on this, we can define the following optimization problem at each CU $q$
\begin{subequations}\label{eq:optProbCU_v2}
	\begin{align}
	\stepcounter{probNum}
	(\mathrm{P\arabic{probNum}})(q) \quad
	\max_{ \mathcal{W}_{q}}\quad & \sum_{u \in \bar{\mathcal{U}}_{q} } \delta_{u}
	\log\left( 1 + 
	\bar{\xi}_{qu}
	\right) 
	\label{eq:eq:optProbCU_v2_obj}
	\\
	\text{s.t.}\quad 
	&
	\sum_{u \in \mathcal{E}_r} \alpha_{ru} \|{\bf w}_{ru}\|_2^2 \le M,
	\quad
	r \in \mathcal{B}_q
	\\
	&
	\sum_{u\in \mathcal{E}_r} \|{\bf w}_{ru}\|_2^2 \le p,
	\quad \quad \quad
	r \in \mathcal{B}_q
	\label{eq:TotalUtilityProblem_U_1_w_eachCU_v2}
	\\
	&
	\bar{\xi}_{qu} =
		\frac{
		{\bf \bar{w}}_{qu}^H {\bf \bar{h}}_{qu,u} {\bf \bar{h}}_{qu,u}^H {\bf \bar{w}}_{qu}
		}
		{
		C_{qu}\left(\mathcal{W}_{q} \right)
		}
	,
	\label{eq:eq:optProbCU_v2_xi}
	\ 
	u \in \bar{\mathcal{U}}_{q}
	\end{align}
\end{subequations}
where the auxiliary variable~\eqref{eq:eq:optProbCU_v2_xi} does not contain the scheduling variables in the set $\mathcal{S}_q$ because the expression includes the beamformers $\mathcal{W}_{q}$ which will be zero for the users not scheduled, thus, the expression of $C_{qu}$ is the same as $B_{qu}$ without including $\mathcal{S}_q$.

Using fractional programming, our optimization problem in~\eqref{eq:optProbCU_v2} can be reformulated as
\begin{subequations}\label{eq:optProbCU_v3}
	\begin{align}
	\stepcounter{probNum}
	(\mathrm{P\arabic{probNum}})(q) \quad
	\max_{ \mathcal{W}_{q}, \bar{\bm \xi}_{q}, \bar{\bm \zeta}_q }\quad & f_2\left( \mathcal{W}_q, \bar{\bm \xi}_{q}, \bar{\bm \zeta}_q \right) & 
	\label{eq:eq:optProbCU_v3_obj}
	\\
	\text{s.t.}\quad 
	&
	\sum_{u \in \mathcal{E}_r} \alpha_{ru} \|{\bf w}_{ru}\|_2^2 \le M,
	\ 
	r \in \mathcal{B}_q
	\label{eq:optProbCU_v3_capacity}
	\\
	&
	\sum_{u\in \mathcal{E}_r} \|{\bf w}_{ru}\|_2^2 \le p,
	\quad \quad
	r \in \mathcal{B}_q .
	\label{eq:optProbCU_v3_power}
	\end{align}
\end{subequations}
with the function~$f_2\left( \mathcal{W}_q, \bar{\bm \xi}_{q}, \bar{\bm \zeta}_q \right)$ in~\eqref{eq:eq:optProbCU_v3_obj} defined as
\begin{align}\label{eq:CUNewObj_v2}
	& f_2\left( \mathcal{W}_q, \bar{\bm \xi}_{q}, \bar{\bm \zeta}_q \right) =
	\sum_{u \in \bar{\mathcal{U}}_{q} }
	\delta_{u} \left( \log\left( 1 + \bar{\xi}_{qu} \right) - \bar{\xi}_{qu} \right)
	\nonumber \\
	& \ 
	+
	\sum_{u \in \bar{\mathcal{U}}_{q} }
	\Bigg(
	2 \text{Re}\left\{
	\bar{\zeta}_{qu}^{*}
	\sqrt{\delta_{u}\left( 1 + \bar{\xi}_{qu}\right)}
	{\bf \bar{w}}_{qu}^H {\bf \bar{h}}_{qu,u}
	\right\}
	\nonumber \\
	& \ 
	-
	|\bar{\zeta}_{qu}|^2
	\left(
	{\bf \bar{w}}_{qu}^H {\bf \bar{h}}_{qu,u} {\bf \bar{h}}_{qu,u}^H {\bf \bar{w}}_{qu}
	+ C_{qu}\left(\mathcal{W}_{q} \right)
	\right)
	\Bigg) ,
\end{align}
where $\bar{\bm \zeta}_{q} = [\bar{\zeta}_{qu_1}\ \dots\  \bar{\zeta}_{qu_{|\mathcal{U}_q|}}]^T$ are required auxiliary variables. In Appendix~\ref{eq:CUsystem_Form1} we provide the details of the reformulation leading to~\eqref{eq:optProbCU_v3}. 

Using the first optimality condition of~\eqref{eq:CUNewObj_v2} with respect to $\bar{\zeta}_{qu}$. The optimal value of $\bar{\zeta}_{qu}$ is
\begin{align}\label{eq:zeta_v2}
\bar{\zeta}_{qu}
=
\frac{
	\sqrt{\delta_{u}\left( 1 + \bar{\xi}_{qu}\right)}
	{\bf \bar{w}}_{qu}^H {\bf \bar{h}}_{qu,u}
}
{
	{\bf \bar{w}}_{qu}^H {\bf \bar{h}}_{qu,u} {\bf \bar{h}}_{qu,u}^H {\bf \bar{w}}_{qu}
	+ C_{qu}\left(\mathcal{W}_{q} \right)
} .
\end{align}

Now, using the constraints~\eqref{eq:optProbCU_v3_capacity} and~\eqref{eq:optProbCU_v3_power}, we define the following Lagrangian function
\begin{align}\label{eq:constraintsLagran}
&f_3(\mathcal{W}_q, {\bm \mu}_q, {\bm \lambda}_q)
=
-
\sum_{r \in \mathcal{B}_q} 
\lambda_r
\left(
\sum_{u \in \mathcal{E}_r} 
\alpha_{ru} \|{\bf w}_{ru}\|_2^2
- M
\right)
\nonumber \\
& \quad
-
\sum_{r \in \mathcal{B}_q} \mu_r \left( 
\sum_{u \in \mathcal{E}_r} \|{\bf w}_{ru}\|_2^2 - p
\right)
\nonumber \\
&
\stackrel{(a)}{=}
-
\sum_{u \in \mathcal{U}_q} \sum_{r \in \mathcal{D}_{qu} }
\lambda_r
\alpha_{ru} \|{\bf w}_{ru}\|_2^2
+ M \sum_{r \in \mathcal{B}_q} \lambda_r
\nonumber \\
& \quad
-
\sum_{u \in \mathcal{U}_q} \sum_{r \in \mathcal{D}_{qu} } \mu_r 
\|{\bf w}_{ru}\|_2^2
+ p \sum_{r \in \mathcal{B}_q} \mu_r
\nonumber \\
&
=
-
\sum_{u \in \mathcal{U}_q} 
\sum_{r \in \mathcal{D}_{qu} }
\omega_{ru} \|{\bf w}_{ru}\|_2^2
+ \sum_{r \in \mathcal{B}_q} \left( p \mu_r + M \lambda_r\right) ,
\end{align}
where $\omega_{ru} = \mu_r + \lambda_r \alpha_{ru}$, and $(a)$ follows from $\sum_{r \in \mathcal{B}_q} \sum_{u \in \mathcal{E}_r} \left(\cdot\right) = \sum_{u \in \mathcal{U}_q} \sum_{ r \in \mathcal{D}_{qu} } \left(\cdot\right)$.
 
When the variables other than $\mathcal{W}_q$ are fixed, using the Lagrangian formulation of~\eqref{eq:optProbCU_v3}, we can write the corresponding expression for the optimal beamformer ${\bf w}_{ru}$ as~\eqref{eq:beamformer_CU_est_v2_concat}, where ${\bm \Omega}_{qu} = \left({\rm diag} \left( \left\{ \omega_{ru} : r \in \mathcal{C}_u \right\} \right) \otimes {\bf I}_M \right)$. Note that the beamformer ${\bf w}_{ru}$ is obtained from ${\bf \bar{w}}_{qu}$ using~\eqref{eq:beamformer_concat_def}.
\begin{figure*}[b]
	\hrule
\begin{align}\label{eq:beamformer_CU_est_v2_concat}
&{\bf \bar{w}}_{qu}
=
\bar{\zeta}_{qu}^{*}
\sqrt{\delta_{u}\left( 1 + \bar{\xi}_{qu}\right)}
\Big(
|\bar{\zeta}_{qu}|^2
\Big(
{\bf \bar{h}}_{qu,u} {\bf \bar{h}}_{qu,u}^H + {\bf \bar{\bm \Theta}}_{qu,u}
+ 
\sum_{u' \in \bar{\mathcal{U}}_{-q}} \bar{t}_{q,u'} \left( {\bf \bar{h}}_{qu,u'} {\bf \bar{h}}_{qu,u'}^H + {\bf \bar{\bm \Theta}}_{qu,u'} \right)
\Big)
\nonumber \\
\noalign{\vskip-7pt}
& \quad \quad \quad \quad \quad \quad \quad \quad \quad
+
\sum_{u' \in \bar{\mathcal{U}}_{q}, u' \ne u}
|\bar{\zeta}_{qu'}|^2
\left( 
{\bf \bar{h}}_{qu,u'} {\bf \bar{h}}_{qu,u'}^H + {\bf \bar{\bm \Theta}}_{qu,u'}
\right)
+ {\bm \Omega}_{qu}
\Big)^{-1}
{\bf \bar{h}}_{qu,u}
\nonumber \\
&\ 
=
\bar{\zeta}_{qu}^{*}
\sqrt{\delta_{u}\left( 1 + \bar{\xi}_{qu}\right)}
\Big(
|\bar{\zeta}_{qu}|^2
\sum_{u' \in \bar{\mathcal{U}}_{-q}} \bar{t}_{q,u'} \left( {\bf \bar{h}}_{qu,u'} {\bf \bar{h}}_{qu,u'}^H + {\bf \bar{\bm \Theta}}_{qu,u'} \right)
+
\sum_{u' \in \bar{\mathcal{U}}_{q}}
|\bar{\zeta}_{qu'}|^2
\left( 
{\bf \bar{h}}_{qu,u'} {\bf \bar{h}}_{qu,u'}^H + {\bf \bar{\bm \Theta}}_{qu,u'}
\right)
+ {\bm \Omega}_{qu}
\Big)^{-1}
{\bf \bar{h}}_{qu,u} ,
\end{align}
\end{figure*}

The Lagrangian multipliers $\mu_r \ge 0$ and $\lambda_r \ge 0$ found in ${\bm \Omega}_{qu}$ and \eqref{eq:constraintsLagran} can be determined using their power budget in~\eqref{eq:optProbCU_v3_power} and capacity~\eqref{eq:optProbCU_v3_capacity} constraints, respectively. Importantly, both constraints are related to the power used at DU $r$, where constraint~\eqref{eq:optProbCU_v3_capacity} can be seen as a weighted power constraint. Hence, both cannot be tight simultaneously; from complementary slackness, one of these Lagrangian multipliers needs to be zero. Unfortunately, we do not know, a priori, which constraint will remain tight. Therefore, we propose a heuristic that, at each iteration, the algorithm checks for whether the capacity constraint is satisfied (allowing $\lambda_r = 0$); if it is not satisfied, we update $\lambda_r$ to a small value. As fo $\mu_r$, we update it using the bisection search method as was discussed in the previous section for a DU-distributed system.

\begin{algorithm}
	\SetAlgoLined
	\SetInd{0.1em}{1em}
	\caption{Distributed resource allocation at each CU $q$}
	\label{algortihm:w_using_weights}
	Initialize $\mathcal{W}$ and $\alpha_{ru}$ for \emph{all} users. \label{step:L_1_norm_weights_init_beams}\\
	\While{ \textbf{NOT} converged}{\label{step:Algo_terminate_2}
		Update $\{ \bar{\xi}_{qu} : u \in \mathcal{U}_q\}$ using~\eqref{eq:eq:optProbCU_v2_xi}.\\
		Update $\{ \bar{\zeta}_{qu}  : u \in \mathcal{U}_q\}$ using~\eqref{eq:zeta_v2}.
		\\
		Update $\mathcal{W}$ using~\eqref{eq:beamformer_CU_est_v2_concat}.\\
		Update $\{\lambda_r, \mu_r : r \in \mathcal{B}_q\}$ as described using complementary slackness.\\
		Update weights ${\bm \alpha}$ using~\eqref{eq:weightsUpdate}.\label{step:using_weights_alpha}\\
	}
\end{algorithm}

It is worth noting that the off-diagonal entries of the terms of the inverse term in~\eqref{eq:beamformer_CU_est_v2_concat} are very small compared to the diagonal entries; this directly follows from summing up a large number of multiplied independent zero-mean random elements. Hence, in practice, changing $\mu_r$ has little effect on the other beamformers $\{ {\bf w}_{r'u}: r' \in \mathcal{D}_{qu}\backslash r \}$ in~\eqref{eq:beamformer_CU_est_v2_concat} (see~\eqref{eq:beamformer_concat_def}).

Finally, we implement Algorithm~\ref{algortihm:w_using_weights} on each CU to perform the resource allocation and the user scheduling. The discussion of Algorithm~\ref{algortihm:w_using_weights} is similar to that in the DU-distributed system and is omitted for brevity.

\section{Scalability of Calculating the Leakage}\label{section:scalability_leakage}
The main issue in calculating the leakage expressions in~\eqref{eq:leakage_DUsystem} or~\eqref{eq:leakage_CUsystem}, is that each DU needs to estimate the channels to all the users in the network. This is, likely, impractical and not scalable. Another issue is that in a distributed algorithm, a DU in the DU-distributed system (or CU in the CU-distributed system) cannot know which users will be scheduled by other DUs (or CUs), i.e., does not know the value of the terms $s_{r,u'}$ (or $\bar{s}_{q,u'}$) to accurately choose $t_{r,u'}$ (or $\bar{t}_{q,u'}$). As such, for a practical implementation, alternative methods to calculate the leakage are needed. 

In this section, we exploit the fact that \textit{leakage is a convenience} we use to estimate the impact a DU (CU) has on other DUs (CUs). We propose here three methods to calculate this impact:
\begin{itemize}
	\item Standard CSI estimation: assume all the users $\mathcal{U}_{-r}$ in DU-distributed system (or $\bar{\mathcal{U}}_{-q}$ in CU-distributed system) are scheduled and estimate the CSI to all of them. This means that for each DU $r$, we set $\{ t_{r,u} = 1 : u \in \mathcal{U}_{-r}\}$; similarly for $\bar{t}_{q,u}$ for the CU-distributed system. This technique is already used above in the development of our algorithms.
	\item Statistical CSI: similar to the first method, except using only the \textit{large-scale fading }statistics to calculate the leakage, eliminating the need for continual training of the leakage channels.
	\item Traffic distribution: use traffic distribution to calculate the leakage. Based on the statistical distribution of traffic, this \textit{completely eliminates }the need for real-time leakage information.
\end{itemize}

\emph{Statistical CSI:} Each DU $r$ uses the large-scale fading statistics to calculate the leakage to \emph{all} the users $u' \in \mathcal{U}_{-r}$ other than the ones it serves. This means that the leakage term in the DU-distributed system can be alternatively defined as
\begin{align}
&L_{ru}\left({\bf w}_{ru}\right) = \sum_{u' \in \mathcal{U}_{-r} }  {\bf w}_{ru}^H {\bm \Lambda}_{ru'} {\bf w}_{ru} ,
\nonumber \\
&\text{where} \quad\quad 
{\bm \Lambda}_{ru'} = \psi_{ru'} \beta(d_{ru'}) {\bf I}_M .
\end{align}
Similarly, for the CU-distributed system, we define the leakage in the CU-distributed system as
\begin{align}
&\bar{L}_{qu}\left( {\bf \bar{w}}_{qu} \right) = \sum_{u' \in \mathcal{U}_{-\bar{\mathcal{U}}_{q}} } {\bf \bar{w}}_{qu}^H {\bar{\bm \Lambda}}_{qu,u'} {\bf \bar{w}}_{qu} ,
\nonumber \\
& \text{where}\ 
{\bar{\bm \Lambda}}_{qu,u'} =
\left({\rm diag} \left( \left\{\psi_{ru'} \beta(d_{ru'}) : r \in \mathcal{D}_{qu} \right\} \right) \otimes {\bf I}_M \right)
.
\end{align}

\emph{Traffic distribution:} Instead of calculating the leakage to all the users that may or may not be scheduled, each node in the network can use a spatial traffic distribution for the network region around it. Such an approach can be easily based on a simple traffic survey, resulting in a traffic probability density function (PDF), $\Upsilon_n(\widetilde{x}_u,\widetilde{y}_u)$ at location $(\widetilde{x}_u,\widetilde{y}_u)$. we can redefine the leakage terms found in~\eqref{eq:SLINR_subterm} and~\eqref{eq:SLINR_subterm_CU_v2} to use the PDF of the traffic distribution in the region around the node instead of calculating the leakage to the users in these regions. Hence, for the DU-distributed system, the leakage term can be written as
\begin{align}\label{eq:leakageTrafficDistribution_DU}
	&L_{ru}\left({\bf w}_{ru}\right)
	=
	\mathbb{E}_{\widetilde{x}_u,\widetilde{y}_u}\left\{
	{\bf w}_{ru}^H
	\tilde{\bm \Lambda}_{ru} {\bf w}_{ru}
	\right\}
	\nonumber \\
	& \quad \quad
	=
	\iint_{\widetilde{x}_u, \widetilde{y}_u \in \iota_{r}}
	{\bf w}_{ru}^H
	\tilde{\bm \Lambda}_{ru} {\bf w}_{ru}
	\Upsilon(\widetilde{x}_u,\widetilde{y}_u)
	\diff{\widetilde{x}_u} \diff{\widetilde{y}_u},
\end{align}
where $\iota_{r}$ is the boundary of the region considered to calculate the leakage. Essentially, \eqref{eq:leakageTrafficDistribution_DU} calculates the average leakage power weighted by the traffic PDF, i.e., it emphasizes the regions with hotspots. The term $\tilde{\bm \Lambda}_{ru} = \left( \beta(\widetilde{d}_{ru} + d_{\rm excl}) \right) {\bf I}_M $, with $\widetilde{d}_{ru} = \sqrt{ \left( \widetilde{x}_r - \widetilde{x}_u \right)^2 + \left( \widetilde{y}_r - \widetilde{y}_u \right)^2 }$ written as a function of the location of the DU $(\widetilde{x}_r,\widetilde{y}_r)$ and the point $(\widetilde{x}_u,\widetilde{y}_u)$ used to calculate the leakage, and $d_{\rm excl}$ is an exclusion region around the DU, which is chosen to be larger than the reference distance of the path loss.

Similarly, in the CU-distributed system, the leakage term can be defined as
\begin{align}
&\bar{L}_{qu}\left( {\bf \bar{w}}_{qu} \right)
=
\mathbb{E}_{\widetilde{x}_u,\widetilde{y}_u}\left\{
{\bf \bar{w}}_{qu}^H
\breve{\bm \Lambda}_{qu} {\bf \bar{w}}_{qu}
\right\}
\nonumber \\
& \quad \quad
=
\iint_{\widetilde{x}_u, \widetilde{y}_u \in \bar{\iota}_{q}}
{\bf \bar{w}}_{qu}^H
\breve{\bm \Lambda}_{qu} {\bf \bar{w}}_{qu}
\Upsilon(\widetilde{x}_u,\widetilde{y}_u)
\diff{\widetilde{x}_u} \diff{\widetilde{y}_u},
\end{align}
where $ \breve{\bm \Lambda}_{qu} = \left({\rm diag} \left( \left\{ \beta\left( \widetilde{d}_{ru} + d_{\rm excl} \right) : r \in \mathcal{D}_{qu}
\right\}
\right) \otimes {\bf I}_M \right)$ is the concatenated path loss between the serving cluster of user $u$ found in cell $q$, i.e., $\mathcal{D}_{qu}$, and locations $(\widetilde{x}_u,\widetilde{y}_u) \in  \bar{\iota}_{q}$. We emphasize that the traffic distribution \textit{eliminates the need for any inter-node information exchange.}

\section{Numerical Results}\label{section:results}
\begin{table}
	\centering
	\begin{tabular}{|p{0.12\linewidth}|p{0.4\linewidth}|p{0.32\linewidth}|}
		\hline
		\hline
		& \multicolumn{1}{l|}{ \textit{\textbf{Parameter}}} & \multicolumn{1}{l|}{\textit{\textbf{Value}}}\\
		\hline
		Cell config. & $Q$, $N$, $M$, density of PPP $\lambda_\text{users}$ when no hotspots exist & $7$, $10$, $8$, $200$ users/$\text{km}^2$\\
		\hline
		Power, pilots & $p$, $p_u$, $\tau_d$, $\tau_p$  & $30~{\rm dBm}$, $20~{\rm dBm}$, $200$, $(32$ or $64)$\\
		\hline
		Noise & noise spectral efficiency $S_z$, noise figure $F_z$, Bandwidth & $-174~{\rm dBm/Hz}$, $8~{\rm dBm}$, $180~{\rm KHz}$\\
		\hline
		Fading &  
		$\sigma_\text{shadowing}$, 
		$\rho$ 
		& 
		$4~{\rm dB}$, 
		$\beta(0.4)$\\
		\hline
		\hline
	\end{tabular}
	\vspace{-0.5em}
	\caption{Simulation parameters.}
	\label{table:sim_parameters}   
	\vspace{-1em}
\end{table}
In this section, we illustrate the efficacy of the proposed algorithms. We simulate a network of $7$ hexagonal \emph{virtual} cells, i.e., $7$ CUs, each has a radius of $500$ meters. We assume that the $N$ DUs in each virtual cell are uniformly distributed inside the cell boundaries. We use wraparound to eliminate the network border effect. As stated in the system model, the user-centric clustering is applied, thus the concept of cells is only applicable on the front-haul, hence the term ``virtual''. We use an exclusion region of radius $20~{\rm m}$ around each DU and model the user locations as a Poisson point process (PPP) with local density determined by the traffic model. 

\begin{figure*}[t]
	\centering
	\begin{subfigure}{0.5\textwidth}
		\centering
		\includegraphics[width=0.96\textwidth]{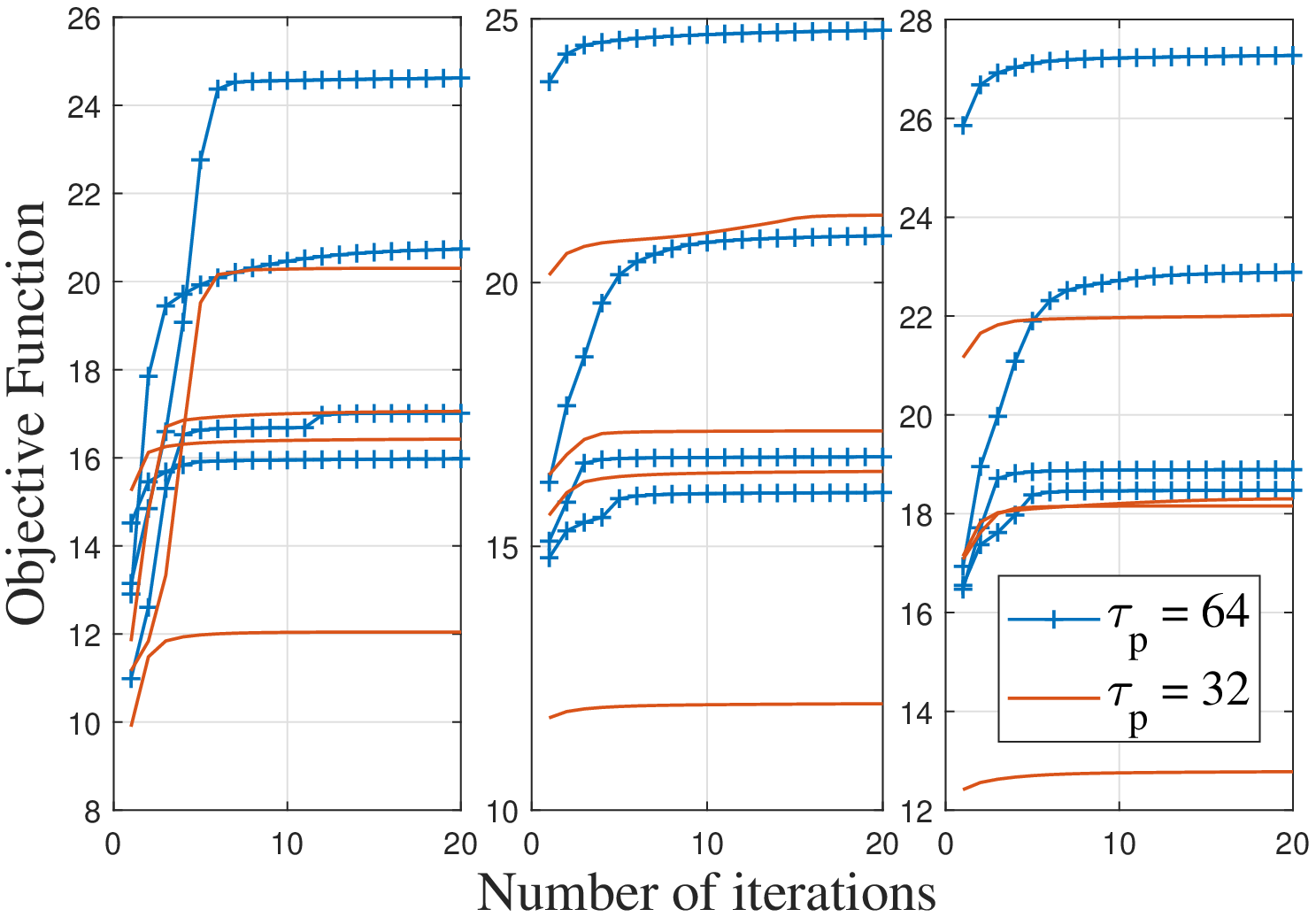}
		\captionof{figure}{On typical DUs in the DU distributed system.}  
		\label{fig:objFunct_singleTS_singleDU_DUsystem}
	\end{subfigure}%
	\begin{subfigure}{0.5\textwidth}
		\centering
		\includegraphics[width=0.96\textwidth]{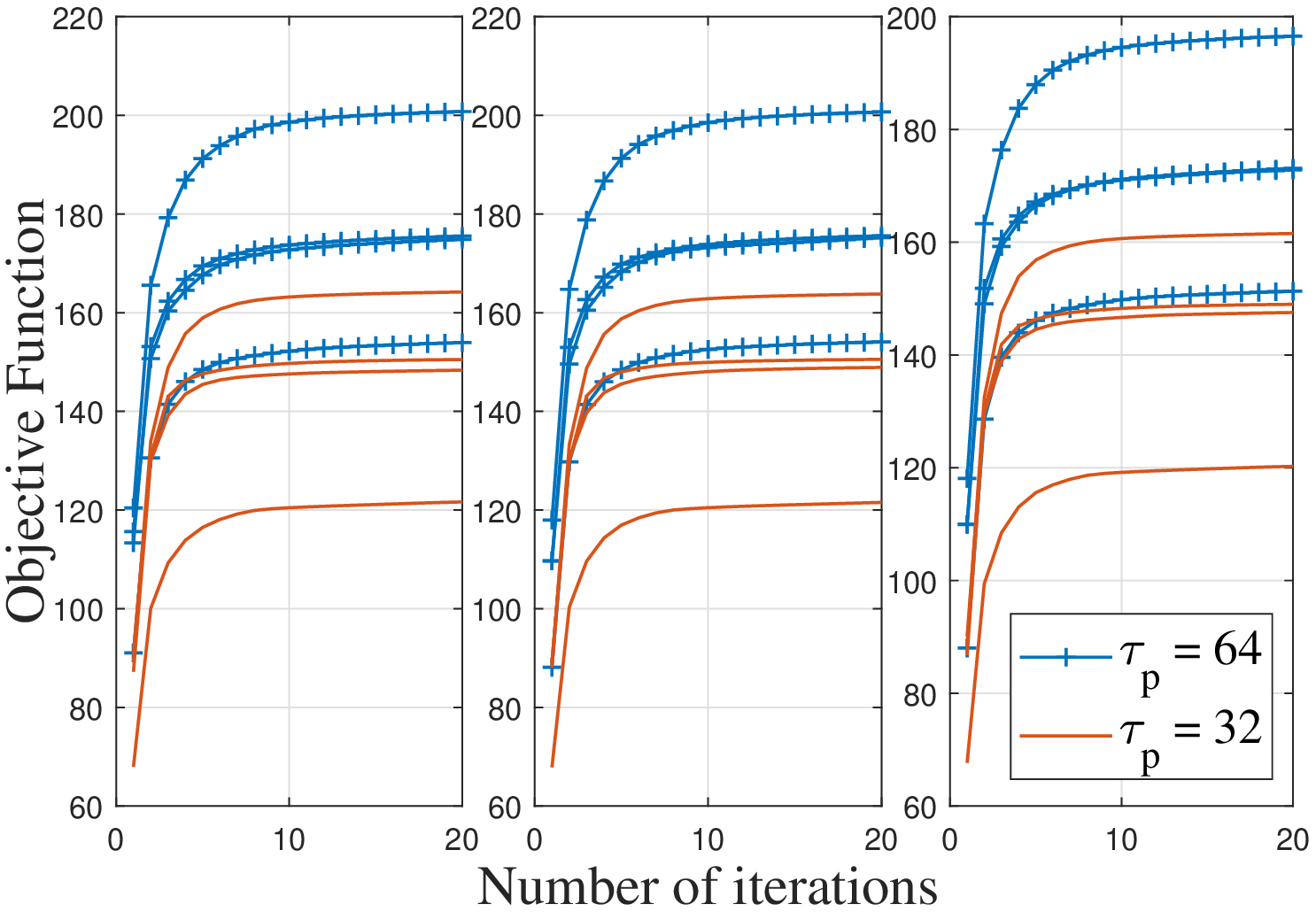}
		\captionof{figure}{On typical CUs in the CU distributed system.} 
		\label{fig:objFunct_singleTS_singleCU_CUsystem}
	\end{subfigure}%
	\caption{Evolution of objection function. Subplots in each system: (Left) standard CSI est. case, (center) statistical CSI for out-of-cluster users case, (right) traffic distribution case.}
	\label{fig:ObjFunctEvolution}
\end{figure*}

In this paper, we model the traffic distribution as a mix of a uniform distribution, chosen with a probability $P_{\rm h}$,  and a number ($N_{\rm h}$) of hotspots. The hotspot traffic is modeled as bivariate normal distributions. Based on this, we define the PDF of the traffic used to calculate the leakage on a node $n$ (DU or CU) as~\cite{Arin8529184}
\begin{align}\label{eq:trafficPDF}
		&\Upsilon_n(\widetilde{x}_u,\widetilde{y}_u) \triangleq f_{\rm h} \left(P_{\rm h} \left( \frac{1}{a_n} \right)
		+ \left( 1 - P_{\rm h} \right)
		\right.
		\nonumber \\
		&\ 
		\left.
		\times
		\frac{1}{N_{\rm h} 2 \pi \sigma_{\rm h}^2} \sum_{i = 1}^{N_{\rm h}} \left( \exp\left(-\frac{\left(\widetilde{x}_k - \widetilde{x}_{{\rm h}_i}\right)^2+\left(\widetilde{y}_k - \widetilde{y}_{{\rm h}_i}\right)^2}{2\sigma_{\rm h}^2}\right) \right) \right)
\end{align}
The value of $P_{\rm h}$ can be used to control the density of the hotspots that are centered at locations $\left[{\bf \widetilde{x}}_{\rm h},{\bf \widetilde{y}}_{\rm h}\right] = \left[\left[{\widetilde{x}}_{h_1},\dots,{\widetilde{x}}_{h _{N_{\rm h}}}\right]^T, \left[{\widetilde{y}}_{h_1}, \dots, {\widetilde{y}}_{h_{N_{\rm h}}}\right]^T\right] \in \mathbb{R}^{N_{\rm h}\times 2}$. These hotspots have equal variances $\sigma_{\rm h}^2$ in both x and y dimensions. The term $a_n$ is the area of the considered region around each node, and $f_{\rm h}$ is a normalizing factor that can be calculated numerically to normalize the PDF.

We simulate path loss using the COST231 Walfisch-Ikegami model~\cite{Walfisch14401}, at carrier frequency $f = 1800$ MHz, in a typical urban environment, where we define $\beta(d_{ru})\left(\mathrm{dB}\right) = - 112.4271 - 38\log_{10}\left(d_{ru}\right)$ with $d_{ru}$ in $\mathrm{km}$. We use $\tau_d = 200$ symbols as the length of the downlink transmission phase, and we simulate the cases of pilot lengths $\tau_p = 32$ and $64$. We average our results using Monte Carlo simulations over both network realizations and time slots. 

Simulating multiple time slots captures the effect of the fairness on user scheduling. In this regard, we simulate $30$ time slots and average the results over the last $20$ time slots after the transient in the fairness weights. These results represent the network steady state performance and is denoted as the long-term performance. In Table~\ref{table:sim_parameters}, we summarize the parameters used in all simulations unless specified otherwise.

In Fig.~\ref{fig:ObjFunctEvolution}, we plot the evolution of the objective function in the DU-distributed system (Fig.~\ref{fig:ObjFunctEvolution}(\subref{fig:objFunct_singleTS_singleDU_DUsystem})) and the CU-distributed one (Fig.~\ref{fig:ObjFunctEvolution}(\subref{fig:objFunct_singleTS_singleCU_CUsystem})) for different realizations of channels and networks. For each system, we show as subplots the three different methods proposed to calculate the leakage. Additionally, we plot many runs for each case different choices of the pilot length, $\tau_p$. As is clear, the results show that the algorithms converge in a smooth, non-decreasing, pattern.

To test the efficacy of our developed approaches and to benchmark them, we compare our results with three schemes defined as follows. 
\begin{itemize}
	\item \label{eq:centralized_SINR} Centralized~\cite{ammarC_RA_UC}: we use a similar approach to the CU-distributed system but with some differences that include the use of the weighted sum rate objective function, i.e., SINR-based. The implication of this is the need for a centralized unit to gather all the information about the network and run the algorithm, which means that the scheme is not as scalable as the distributed schemes.

	In theory, a centralized solution should outperform a distributed solution. However, since the optimization problem is non-convex, any solution is a local optimum. It is, therefore, not guaranteed that the centralized solution outperforms the distributed solution \emph{in practice}.

	\item Local zero-forcing (ZF): we use ZF constructed locally on each DU with round-robin scheduling for the users. Note that the ZF matrix needs to be constructed locally at the DUs, because the sets of users to be served by the DUs overlap with each other. Hence, no disjoint regions can be defined to construct a ZF matrix across multiple DUs. 
	
	\item Conjugate: we use conjugate beamforming with round-robin scheduling.
\end{itemize}

\begin{figure}
	\centering
	\begin{subfigure}{0.48\textwidth}
		\centering
		\includegraphics[width=1\linewidth]{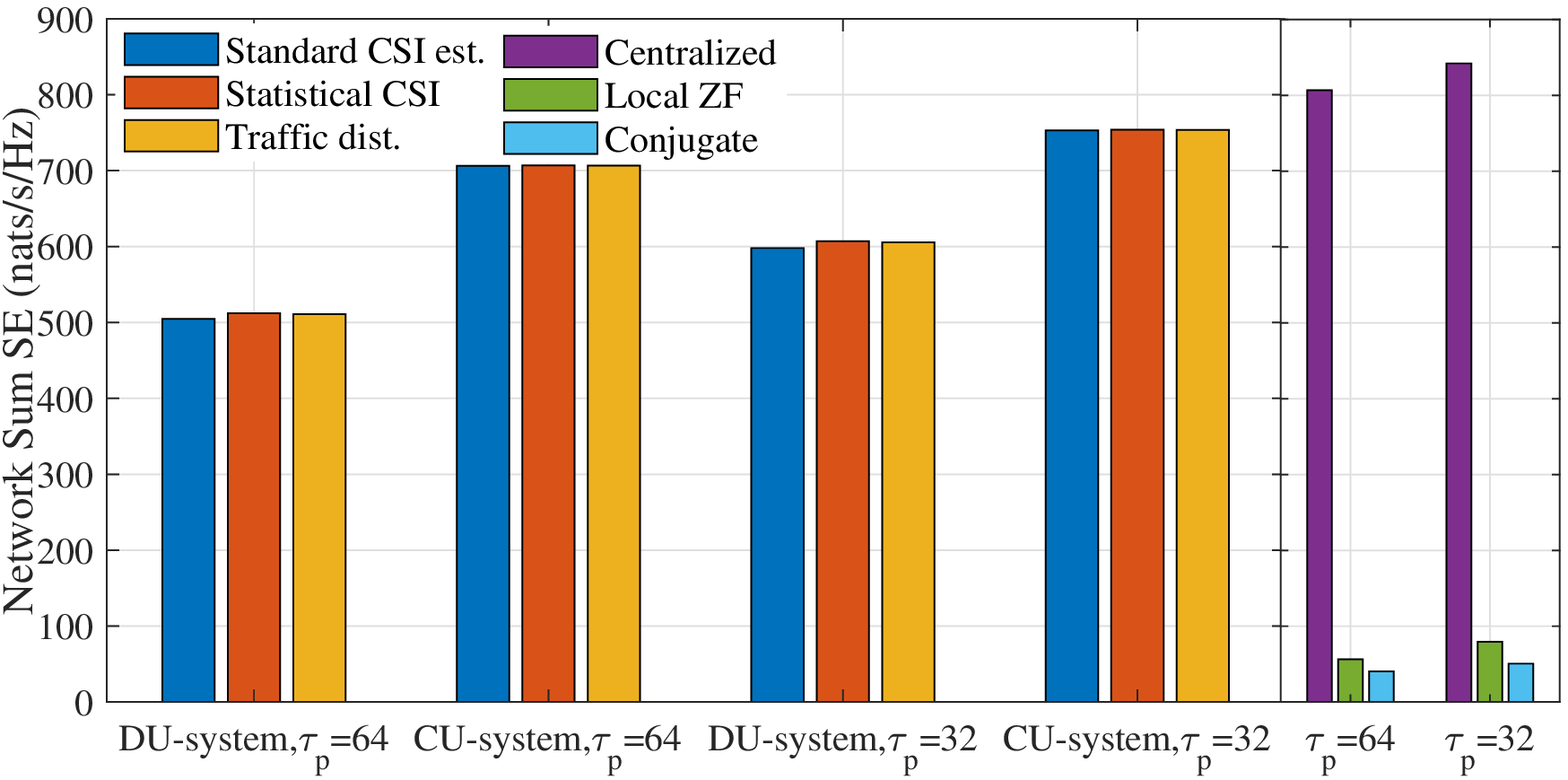}
		\caption{At equal fairness weights (max sum rate).}
		\label{fig:NetSumSE_singleTS}
	\end{subfigure}
	\\
	\begin{subfigure}{0.48\textwidth}
		\centering
		\includegraphics[width=1\linewidth]{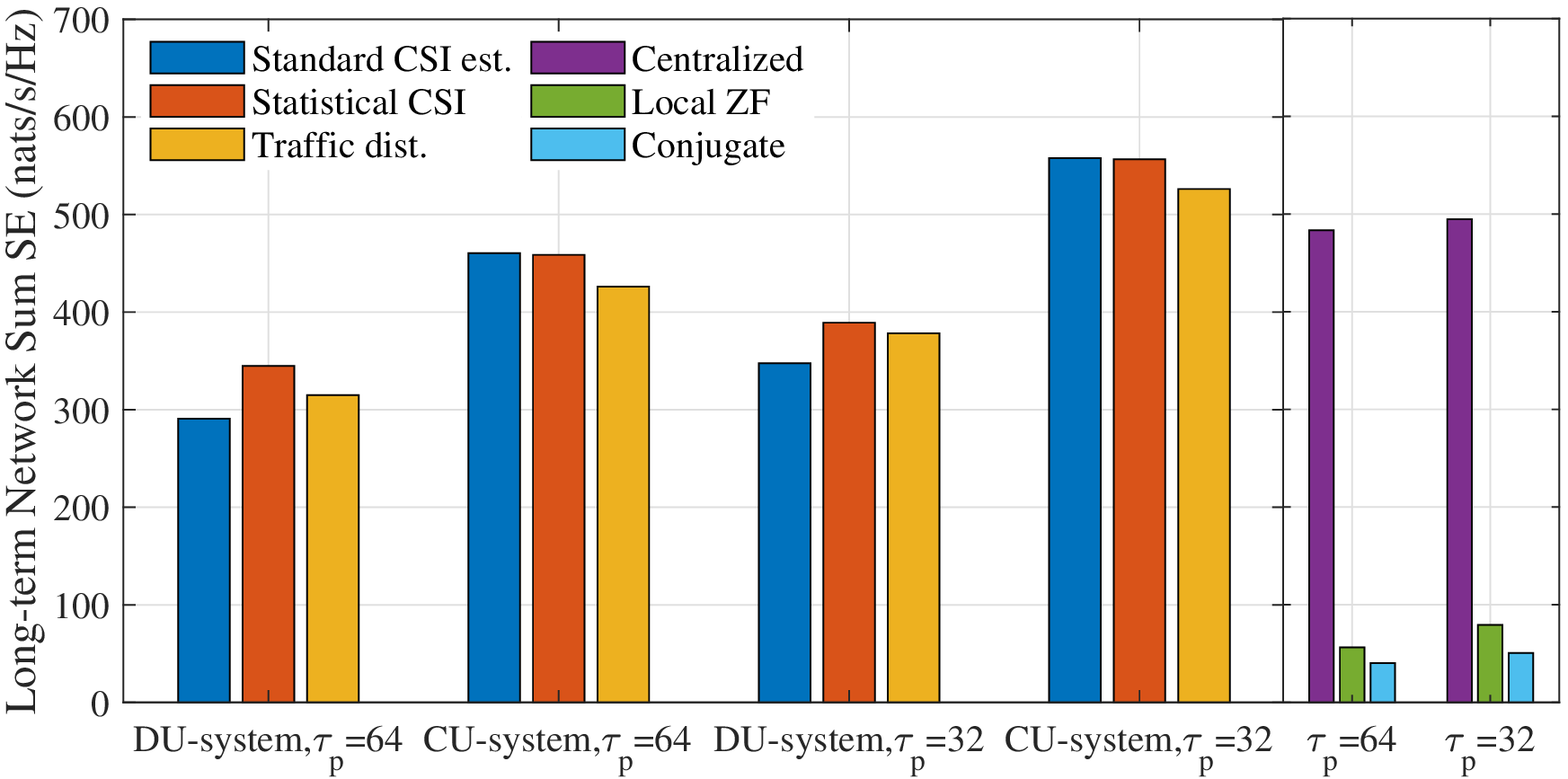}
		\caption{Long-term with evolving fairness weights.}
		\label{fig:NetSumSE_longTerm}
	\end{subfigure}
	\caption{Network sum SE for the two distributed systems under different techniques to calculate leakage.}
	\label{fig:NetSumSE_Uniform}
\end{figure}

\textit{Sum Rate}: In Fig.~\ref{fig:NetSumSE_Uniform}(\subref{fig:NetSumSE_singleTS}), we plot the network sum of spectral efficiency (SE) averaged over network realizations for a single time slot when the fairness weights are equal for all the users (max sum rate). As expected, the CU-distributed system provides better performance than the DU-distributed system because it implements more coordination. Compared to a centralized approach, the CU-system provides comparable performance, with only a slight loss in network sum SE. This is a very important result because it illustrates the importance of deploying multiple CUs in the cell-free MIMO scheme. Note that unlike the DU- and CU-distributed systems, the centralized scheme is not scalable because it requires gathering all the CSI, including inter-CU CSI, at a central node in the network to perform the optimization.

Clearly, round-robin scheduling and the benchmark beamforming systems provide poor performance due to scheduling the users irrespective of their channel conditions and without any consideration of interference cancellation. Compared to the benchmark schemes, our proposed approaches provide a huge gain. Moreover, the results show the efficacy of the different methods of calculating the leakage, denoted as standard CSI, statistical CSI and traffic distribution; these three approaches provide approximately equal performance. One note though is that using the traffic distribution to calculate the leakage is more reasonable when the number of users requesting access is very large, which is the main theme for this study. When the number of users requesting access is small, using the traffic distribution will lead to some loss in performance. 

In Table~\ref{table:comparisonPerformance}, we compare the gains of our proposed schemes to the benchmark schemes, where our two distributed systems are able to provide different degrees of performance. Note that the CU-distributed system provides a $1.28$- and $1.77$-fold performance gain compared to the DU-distributed system using $\tau_p = 32$ and $\tau_p = 64$, respectively.


\textit{Including Fairness}: In Fig.~\ref{fig:NetSumSE_Uniform}(\subref{fig:NetSumSE_longTerm}), we plot the long-term network sum SE, averaged over both network realizations and time slots. The importance of this figure is in capturing the effect of the evolution of fairness weights and steady state algorithm performance. The results show that on the long-term, using the statistical CSI provides the best performance among the other methods used to calculate the leakage. Also, the CU-distributed system provides a \CUGainComparedToDULongTerm-fold performance gain over the long-term compared to the DU-distributed system. Furthermore, the results show that over the long-term the CU-distributed system can outperform the centralized scheme. 

\begin{table}[t]
	\centering
	\begin{tabular}{|p{0.14\linewidth}|p{0.15\linewidth}|p{0.15\linewidth}|p{0.15\linewidth}|p{0.15\linewidth}|}
		\hline
		\hline
		& \multicolumn{2}{c|}{DU-distributed system} & \multicolumn{2}{c|}{CU-distributed system} \\
		\cline{2-5}
		& \textit{\textbf{$\tau_p = 64$}} & \textit{\textbf{$\tau_p = 32$}} & \textit{\textbf{$\tau_p = 64$}} & \textit{\textbf{$\tau_p = 32$}}
		\\
		\hline
		Centralized	& $0.64$-fold & $0.72$-fold & $0.88$-fold & $0.9$-fold \\
		\hline
		Local ZF & $9$-fold & $7.6$-fold & $12.5$-fold & $9.5$-fold \\
		\hline
		Conjugate & $12.7$-fold & $12$-fold & $17.5$-fold & $15$-fold \\
		\hline
		\hline
	\end{tabular}
	\caption{Performance of our proposed systems compared to different schemes at equal fairness weights.}
	\label{table:comparisonPerformance}   
	\vspace{-1em}
\end{table}

As mentioned, both the centralized scheme and the ones proposed here result in local optima; given the complexity of the objective function and constraints, it is impossible to bound the distance from the optimal solution. However, as in the proposed distributed schemes, fewer variables are optimized at each DU/CU. We speculate that the fewer variables required for the CU-distributed system lead to a solution closer to the global optimum. This is consistent with results for a traditional cellular network~\cite{9199126}. Another possible reason for this is that the fairness weights of the users are different in each system, after the allocation of time slots, which may affect the set of users being scheduled concurrently in the latter time slots. For the DU-distributed system, the centralized scheme still provides a better performance in any scenario. This is also reasonable because the DU-distributed sytem restricts its performance to the knowledge found at the DU, so it is less cooperative than the CU-distributed system.

All in all, the results show that both schemes are very efficient in performing the resource allocation, though, as expected, the CU-distributed system provides substantially better performance. Our results highlight the importance of deploying multiple CUs in the user-centric cell-free network while the DU-distributed system can still be used if the front-haul is overloaded.


\begin{figure*}
	\begin{minipage}{0.48\textwidth}
		\centering
		\vspace{0.2em}
		\includegraphics[width=1\linewidth]{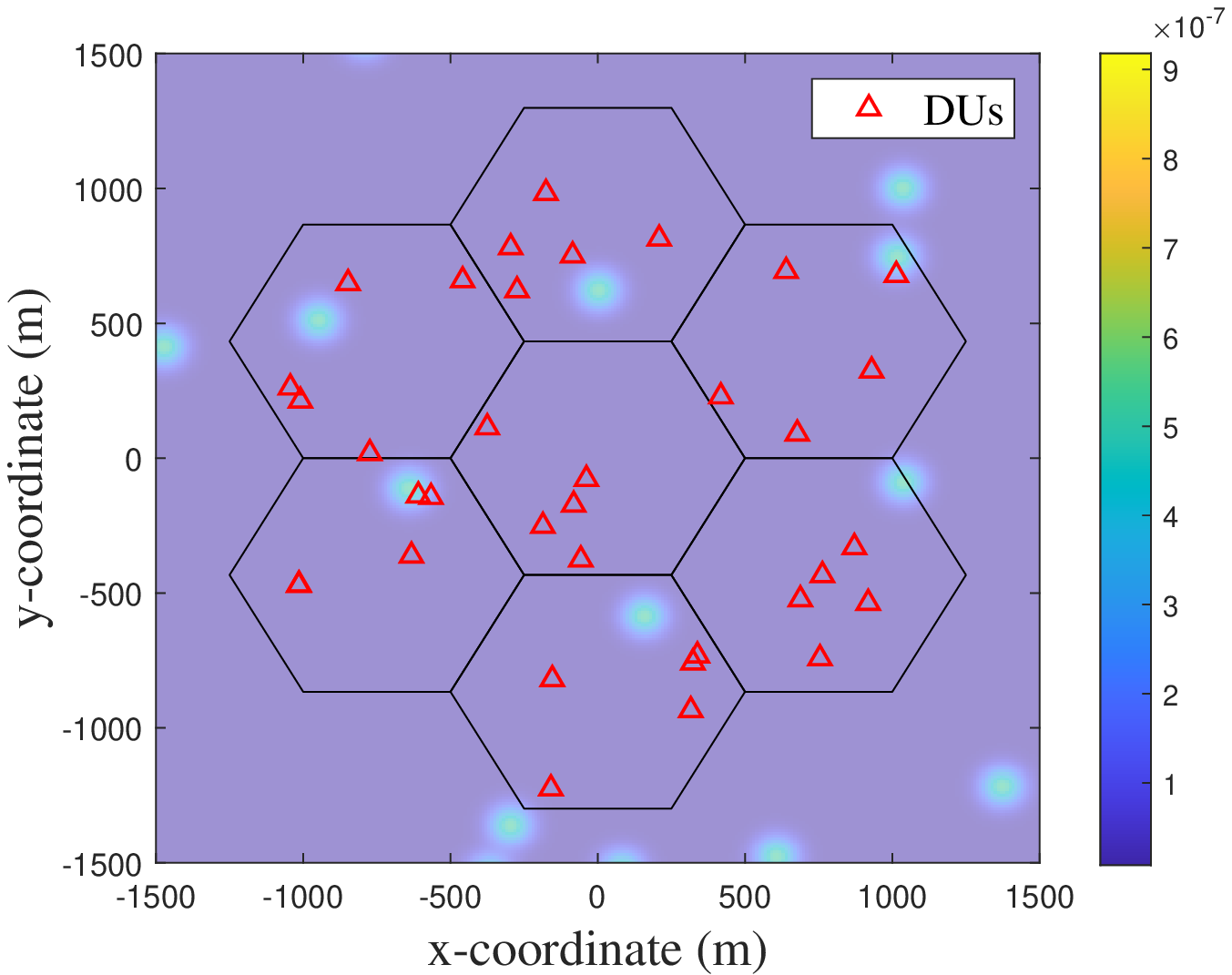}
		\caption{A typical network with hotspot areas, $N=5$, $\lambda_\text{users} = 400$ ${\rm users}/{\rm km}^2$.}
		\label{fig:TypicalHotspostsNet_N5_lambdaUsers400}
	\end{minipage}
	$\ \ \ $
	\begin{minipage}{0.48\textwidth}
		\centering
		\includegraphics[width=1\linewidth]{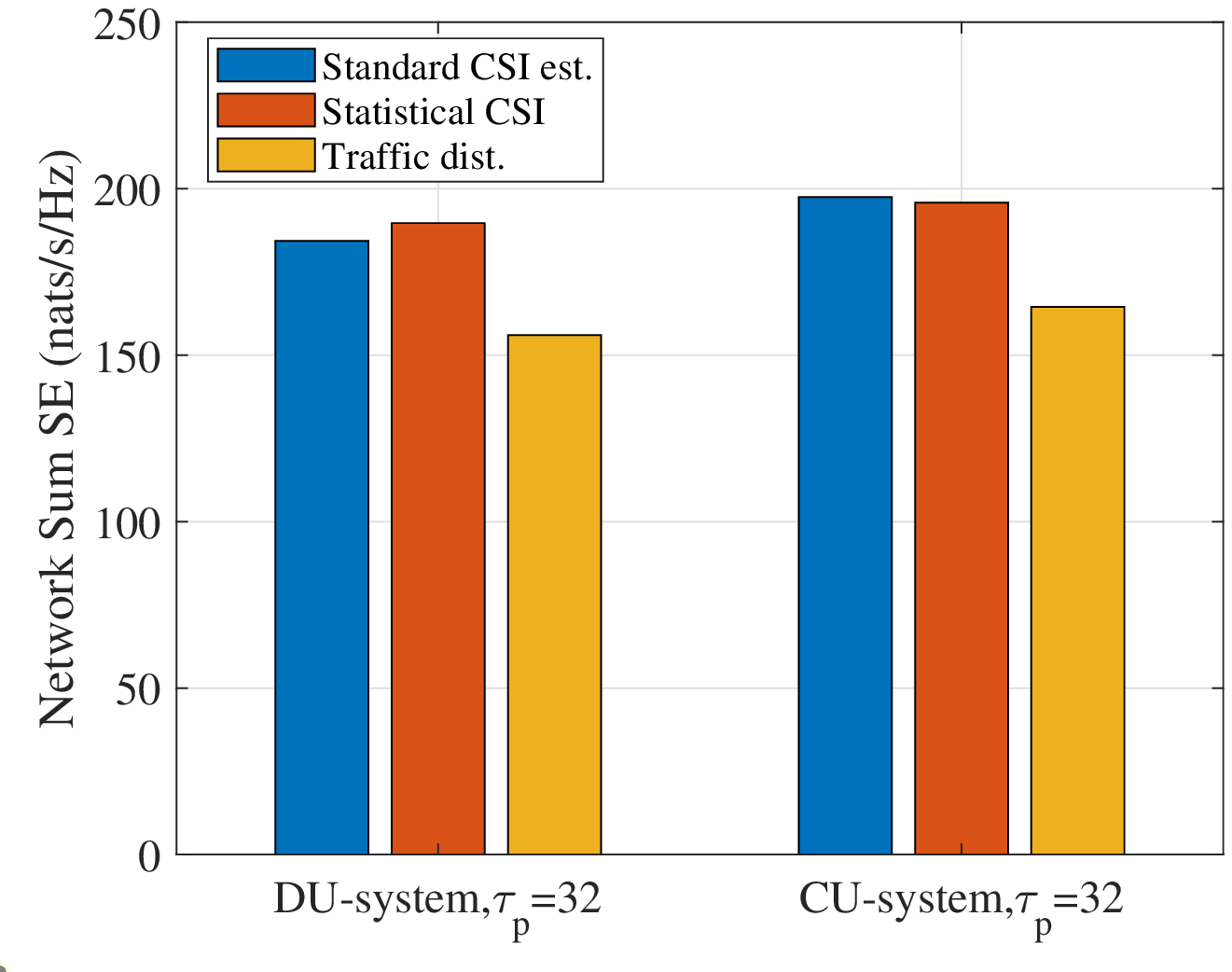}
		\caption{Long-term network sum SE, hotspots scenario, $N=5$, $\lambda_\text{users} = 400$ ${\rm users}/{\rm km}^2$.}
		\label{fig:NetSumSE_fromTS1tillTS10_hotspots}
	\end{minipage}
\end{figure*}

\textit{Using Traffic Distribution}: We use the traffic distribution defined in~\eqref{eq:trafficPDF} with $\sigma_{\rm h}=50~\rm{m}$, $P_{\rm h}=0.5$, and the number of hotspots $N_{\rm h}$ randomly generated as uniformly random variables between four and six hotspots within the studied network area. User locations are then generated from this traffic distribution. An example of a typical network is shown in Fig.~\ref{fig:TypicalHotspostsNet_N5_lambdaUsers400} which also plots a realization of DU locations. In Fig.~\ref{fig:NetSumSE_fromTS1tillTS10_hotspots}, we plot the long-term network sum SE under the scenario of hotspots, where $N=5$ and density of users is $\lambda_\text{users} = 400$ ${\rm users}/{\rm km}^2$. As before, the results are averaged over network realizations and time slots. Most importantly, the results show that the case using the traffic distribution to calculate the leakage provides comparable performance to the other two methods. We conclude that using the traffic distribution to calculate the leakage can be a promising approach to completely eliminating any information exchange between DUs/CUs.

\begin{figure*}[t]
		\begin{minipage}{1\textwidth}
		\centering
		\begin{subfigure}{0.32\textwidth}
			\centering
			\includegraphics[width=1\textwidth]{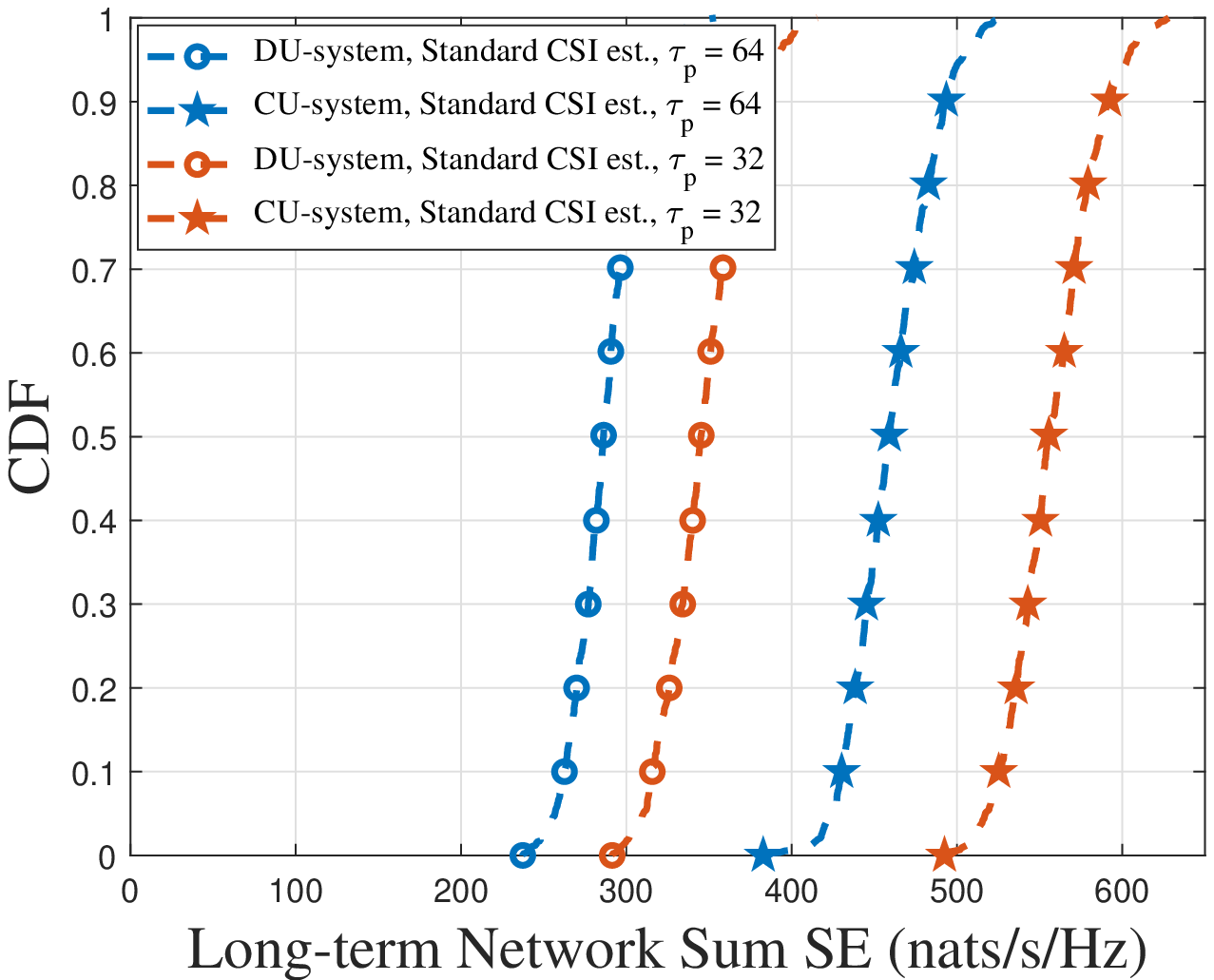}
			\captionof{figure}{Using standard CSI estimation.}
			\label{fig:CDF_NetSumSE_standardCSI}
		\end{subfigure}
		\begin{subfigure}{0.32\textwidth}
			\centering
			\includegraphics[width=1\textwidth]{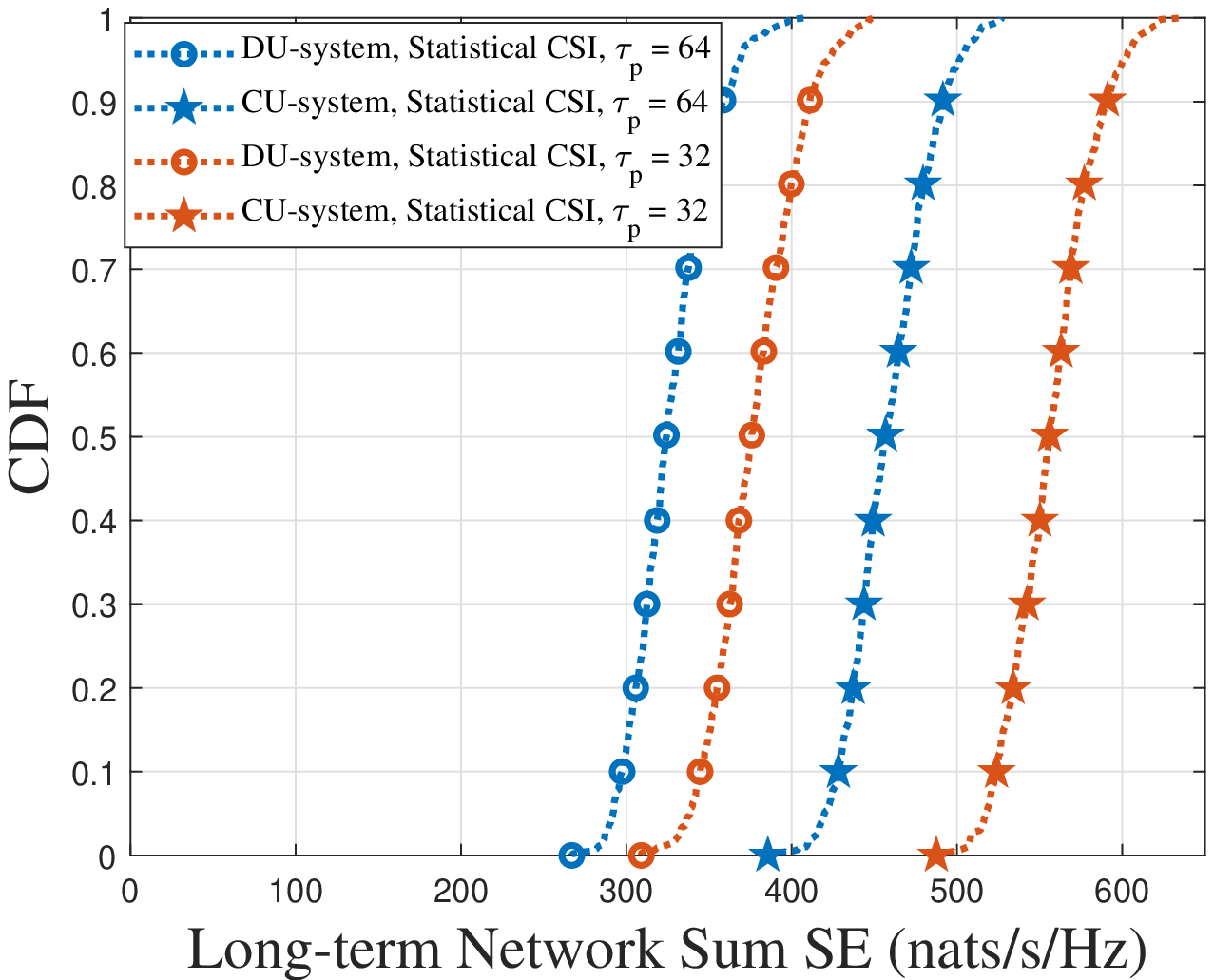}
			\captionof{figure}{Using statistical CSI.}
			\label{fig:CDF_NetSumSE_statisticalCSI}
		\end{subfigure}
		\begin{subfigure}{0.32\textwidth}
			\centering
			\includegraphics[width=1\textwidth]{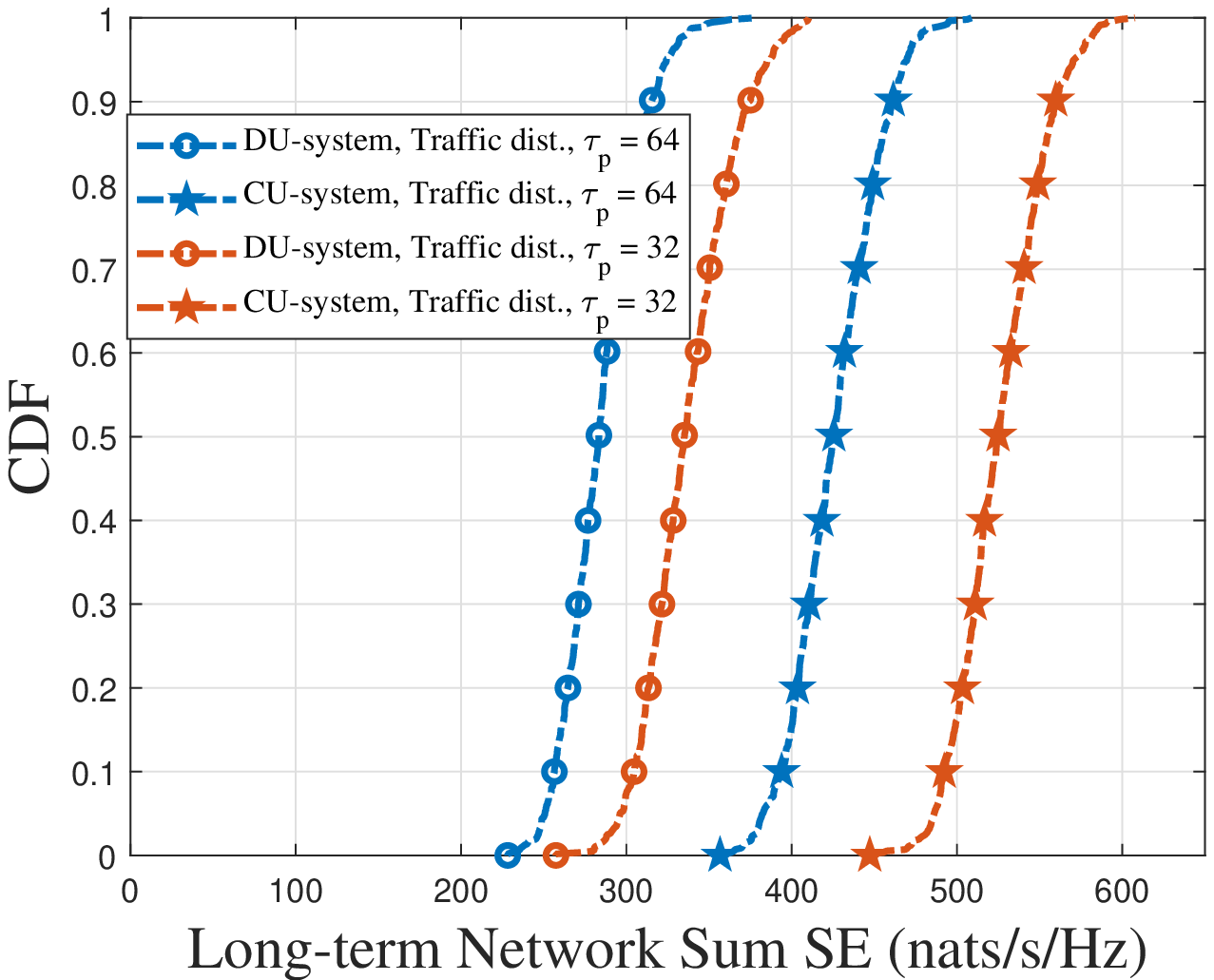}
			\captionof{figure}{Using traffic distribution.}
			\label{fig:CDF_NetSumSE_dist}
		\end{subfigure}
		\caption{CDF of the network sum of spectral efficiency averaged over all the time slots.}
		\label{fig:CDF_NetSumSE_allTSs}
	\end{minipage}
	\\
	\begin{minipage}{1\textwidth}
	\centering
	\begin{subfigure}{0.32\textwidth}
		\centering
		\includegraphics[width=1\textwidth]{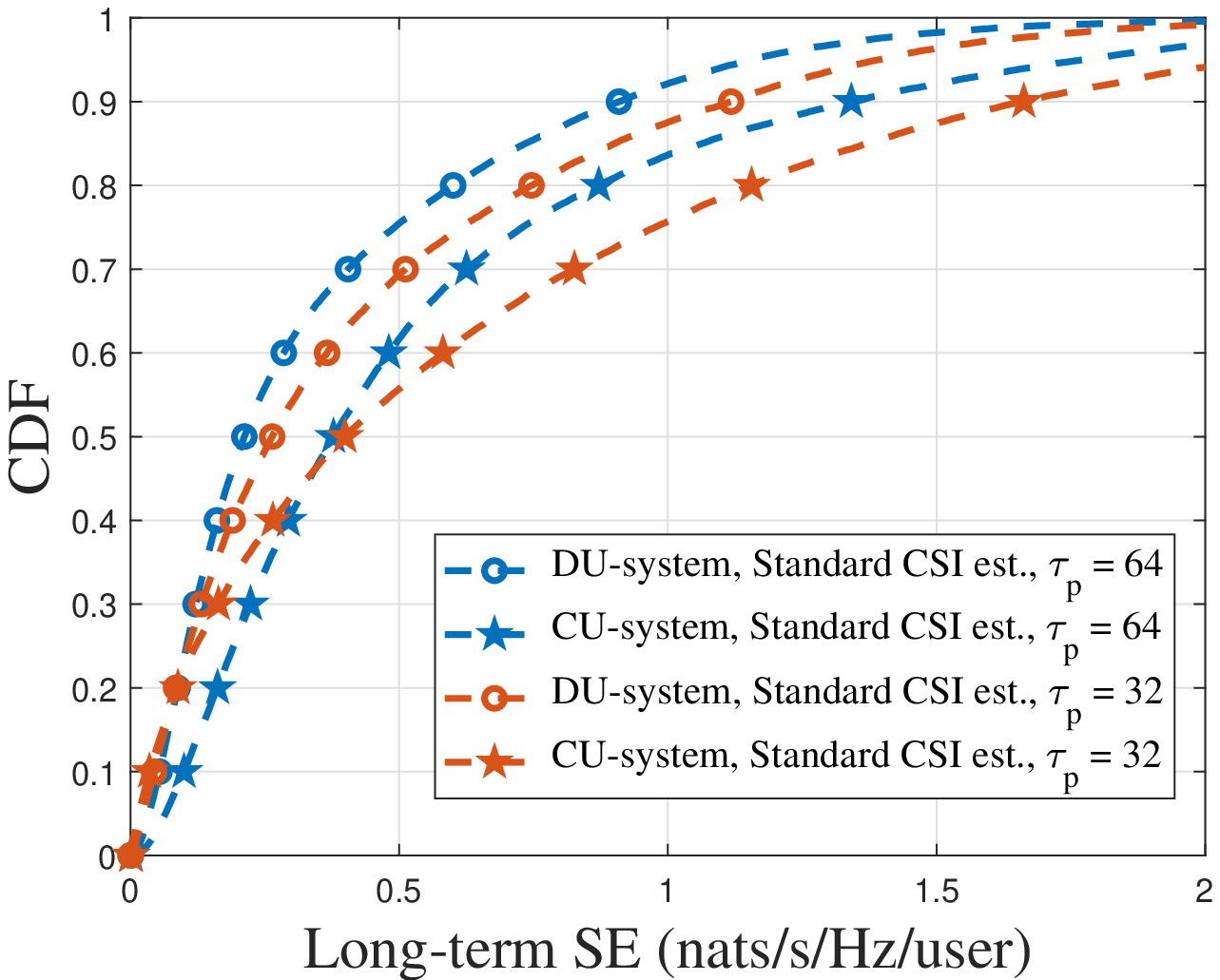}
		\captionof{figure}{Using standard CSI estimation.}
		\label{fig:CDF_allUser_standardCSI}
	\end{subfigure}
	\begin{subfigure}{0.32\textwidth}
		\centering
		\includegraphics[width=1\textwidth]{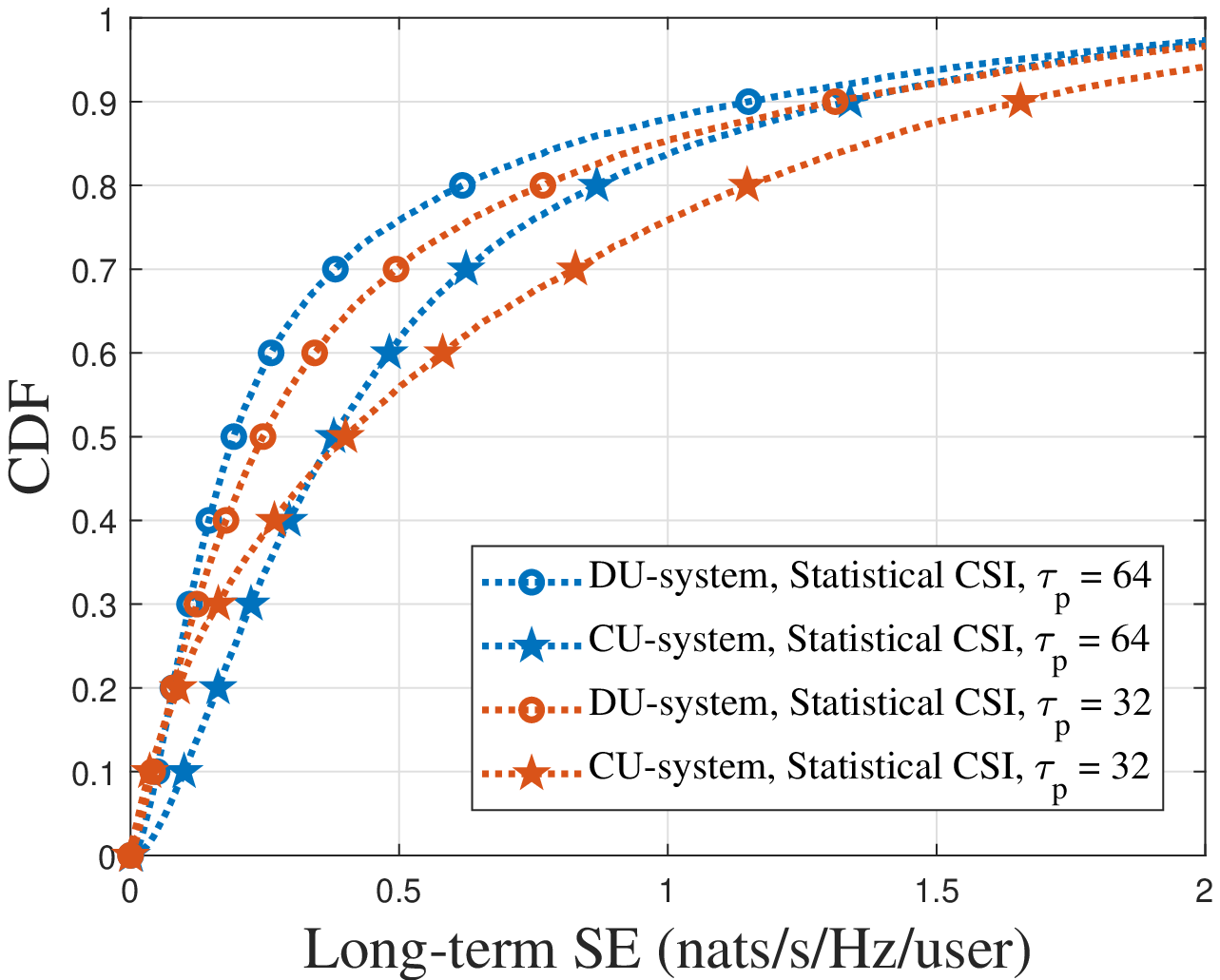}
		\captionof{figure}{Using statistical CSI.}
		\label{fig:CDF_allUse_statisticalCSI}
	\end{subfigure}
	\begin{subfigure}{0.32\textwidth}
		\centering
		\includegraphics[width=1\textwidth]{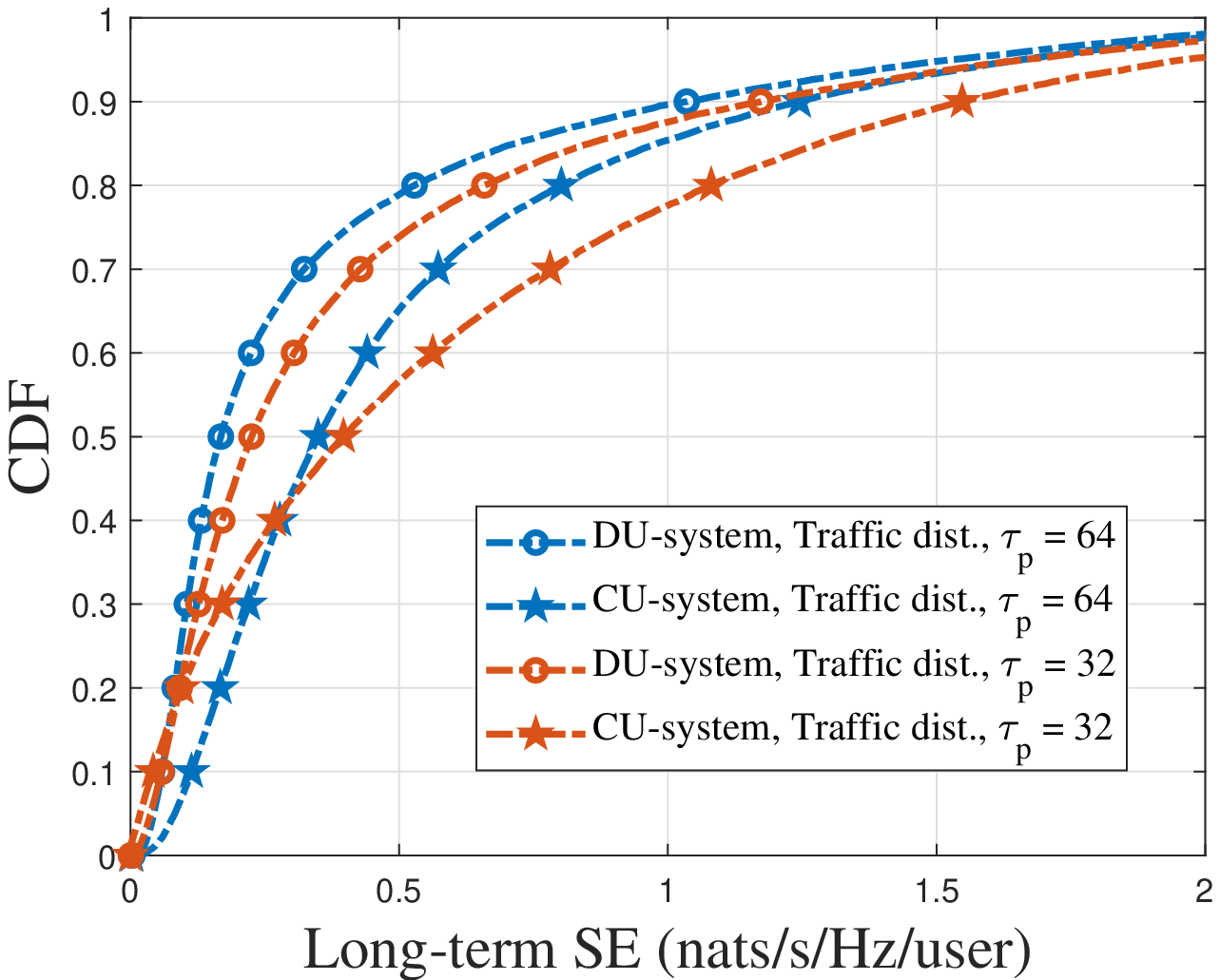}
		\captionof{figure}{Using traffic distribution.}
		\label{fig:CDF_allUser_dist}
	\end{subfigure}
	\caption{CDF of the user spectral efficiency averaged over all the time slots.}
	\label{fig:CDF_Allusers_allTSs}
	\end{minipage}
\end{figure*}

\textit{CDF}: In Figs.~\ref{fig:CDF_NetSumSE_allTSs} and~\ref{fig:CDF_Allusers_allTSs}, we plot the cumulative distribution function (CDF) of the long-term network sum SE and the long-term SE per user, respectively. Note that for the SE per user, we average the SE over all the simulated time slots even if the user is not scheduled; this provides a true measure of the user's overall throughput. The results show the superiority of the CU-distributed system over the DU-distributed system in both network sum SE and SE per user. Notably, for the scenario considered, a pilot length of $\tau_p = 32$ provides better a trade-off between pilot contamination and pilot training overhead compared to~$\tau_p = 64$.

\emph{Algorithm Complexity Analysis}: We now discuss the complexity of the algorithms developed. For Algorithm~\ref{algorithm:SLNR_DU} executed on each DU, per iteration, we have complexity of $\mathcal{O}\left( |\mathcal{E}_r| \right)$ and $\mathcal{O}\left( |\mathcal{E}_r| \right)$ to update ${\bm \xi}_r$ and ${\bm \zeta}_r$, respectively. The combinatorial search step using the Hungarian algorithm has a polynomial time of complexity $\mathcal{O}\left( |\mathcal{E}_r|^3 \right)$. To finish, the complexity of the beamforming step is equal to the complexity of the weighted MMSE which can be derived to be $\mathcal{O}\left( |\mathcal{U}_{\rm s}|^2 M^2 + |\mathcal{U}_{\rm s}| M^3 \right)$~\cite{WMMSE5756489}, where $|\mathcal{U}_{\rm s}| \le M$ is the number of scheduled users. This means that the complexity per iteration of the algorithm is at most $\mathcal{O}\left( M^4 + |\mathcal{E}_r|^3 \right)$ at each DU. 

For Algorithm~\ref{algortihm:w_using_weights} executed on each CU, we have a complexity of $\mathcal{O}\left( |\mathcal{U}_q| \right)$, $\mathcal{O}\left( |\mathcal{U}_q| \right)$, $\mathcal{O}\left( N |\mathcal{E}_{\rm avg}| \right)$, $\mathcal{O}\left( N \right)$ to update $\bar{\bm \xi}_q$, $\bar{\bm \zeta}_q$, ${\bm \alpha}$, and $\{\lambda_r : r \in \mathcal{B}_q\}$ respectively, where $|\mathcal{E}_{\rm avg}|$ is the average number of users to be served by a DU. To compare both systems, we can set $\mathcal{E}_r = \mathcal{E}_{\rm avg}$ in the DU-distributed system. Note that $\mathcal{U}_q \ll N \mathcal{E}_r$ due to the users overlapped between the clusters $\{\mathcal{E}_r : r \in \mathcal{B}_q\}$. The complexity of the beamforming step is at most $\mathcal{O}\left( N^2M^4 + NM^4 \right)$ assuming $|\mathcal{U}_{\rm s}| \le NM$. Hence, we obtain a total complexity of $\mathcal{O}\left( N^2M^4 + N |\mathcal{E}_{\rm avg}| + |\mathcal{U}_q|\right)$ per~CU.

%
As a point of comparison, if we are to construct a centralized resource allocation scheme, we end up with complexity of $\mathcal{O}\left( |\mathcal{U}| \right)$, $\mathcal{O}\left( |\mathcal{U}| \right)$, $\mathcal{O}\left( QN |\mathcal{E}_{\rm avg}| \right)$, $\mathcal{O}\left( QN \right)$ to update $\bar{\bm \xi}$, $\bar{\bm \zeta}$, ${\bm \alpha}$, and $\{\lambda_r : r \in \mathcal{B}\}$, respectively. Hence, we end up with a total complexity per iteration as \sloppy{$\mathcal{O}\left( Q^2 N^2 M^4 + QN |\mathcal{E}_{\rm avg}| + |\mathcal{U}| \right)$} on the centralized node that will run the algorithm. Importantly, such a centralized system requires rapidly increasing computation capabilities in the sense that when the network area grows, the number of virtual cells $Q$ grows. The distributed cases do not suffer from this issue.

From a communication point of view, in a centralized system, all the DUs and the CUs in the network need to communicate the algorithm data with a centralized node at the network core to be able to perform the allocation. All in all, the DU-distributed system is the most distributed scheme, while the CU-distributed system provides a balance between the DU-distributed system and the centralized one in both complexity, scalability and performance.

\begin{table*}[t]
	\centering
	\begin{tabular}{|p{0.2\linewidth}|p{0.28\linewidth}|p{0.15\linewidth}|p{0.1\linewidth}|p{0.1\linewidth}|}
		\hline
		\hline
		\textit{\textbf{System}} & \textit{\textbf{Complexity per node and iteration}} & \multicolumn{3}{c|}{ \textit{\textbf{Front-haul overhead for algorithm}} }
		\\
		\cline{3-5}
		& & {\textit{Standard CSI est.}} & {\textit{Statistical CSI}} & {\textit{Traffic dist.}}
		\\
		\hline
		DU-distributed system & $\mathcal{O}\left( M^4 + |\mathcal{E}_{\rm avg}|^3 \right)$ & 0 & 0 & 0 \\
		\hline
		CU-distributed system & $\mathcal{O}\left( N^2M^4 + N |\mathcal{E}_{\rm avg}| + |\mathcal{U}_q| \right)$ & $\mathcal{O}\left( |\mathcal{U}_{\rm ng}| + N|\mathcal{E}_{\rm avg}| \right)$ & $\mathcal{O}\left( N|\mathcal{E}_{\rm avg} | \right)$ & $\mathcal{O}\left( N|\mathcal{E}_{\rm avg} | \right)$ \\
		\hline
		Centralized & $\mathcal{O}\left( Q^2N^2M^4 + QN |\mathcal{E}_{\rm avg}| + |\mathcal{U}| \right)$ & \multicolumn{3}{c|}{ $\mathcal{O}\left( Q|\mathcal{U}_{\rm ng}| + Q N|\mathcal{E}_{\rm avg}| \right)$} \\
		\hline
		\hline
	\end{tabular}
	\vspace{-0.5em}
	\caption{Comparison of \emph{algorithm complexity and front-haul overhead} for different systems.}
	\label{table:comparisonComplexity}
\end{table*}

\emph{Front-haul Footprint:} To analyze the front-haul communications imposed by the algorithms, the DU-distributed system does not require exchanging information with the CU during the execution of the algorithm to perform the resource allocation. For the CU-distributed system, the $N$ DUs need to exchange, with their CU, the CSI for their users, and then receive the scheduling decisions for $|\mathcal{E}_{\rm avg}|$ users, and the constructed beamformers for at most $M$ users that will be scheduled.

Regarding the CSI exchange, for the standard CSI estimation method used to calculate the leakage, each DU needs to exchange the CSI for $|\mathcal{U}_{\rm ng}|$ users with the CU, where $\mathcal{U}_{\rm ng}$ represents the users within an area around the DU that have a non-negligible reachable signal, which can simply be based on distance or selected regions. For the statistical CSI approach used to calculate the leakage, we need to exchange $|\mathcal{E}_{\rm avg}| \le |\mathcal{U}_{\rm ng}|$ CSI vectors for the users. For the approach that uses the traffic distribution, we do not need to exchange the leakage CSI, because we use a surveyed spatial traffic distribution to calculate the leakage; however the DUs still need to exchange $|\mathcal{E}_{\rm avg}| \le |\mathcal{U}_{\rm ng}|$ CSI vectors for their users. Note that the size of these data to be exchanged can be reduced using signal quantization techniques~\cite{maxMinRate8756286, EnergyEfficiency8781848} proposed for the front-haul. In Table~\ref{table:comparisonComplexity}, we compare the complexity and front-haul footprint of our distributed systems with a centralized one.

\newcommand{\varHowToChoose}{Finally, whether to implement the DU-distributed or the CU-distributed system is a system preference, and it depends on the tradeoff between performance and both computational complexity and fronthaul load. However, we believe that the CU-distributed system (in a multi-CU network) is better because it provides a tradeoff between complexity per node and performance compared to the centralized scheme and the DU-distributed system. Moreover, it allows simple design for the DUs.}
\label{page:HowToChoose}\varHowToChoose

\section{Conclusion}\label{section:conclusion}
In this paper, we developed two distributed algorithms that perform user scheduling, beamforming, and implicit power control for user-centric cell-free MIMO networks. The first system is implemented at the DUs, while the second one is implemented at the CUs. Both use a hybrid metric that depends on the leakage and intra-node interference allowing a distributed allocation. Additionally, exploiting the fact that the  notion of leakage is a tool to measure a transmission has on neighboring ``cells'', we proposed three approaches to calculate the leakage, each requiring a different level of CSI exchange. We present possible deployment scenarios for user-centric cell-free network with manageable complexity with respect to the network size.

In terms of performance, the CU-distributed system provides \CUGainComparedToDULongTerm- to \CUGainComparedToDUInstantanousRounded-fold network throughput gain compared to the DU-distributed system, with a slight increase in complexity and front-haul load - and can outperform centralized solutions. Further, our results measure the performance gain, computational complexity, and front-haul overhead compared to benchmark schemes and a centralized resource allocation. Finally, by analyzing the trade-offs provided by the CU-distributed system compared to DU-distributed and centralized ones, we highlight the importance of deploying multiple CUs in user-centric cell-free networks.

\appendices
\section{Problem reformulation for DU-distributed system}
\label{eq:DUsystem_Form1}
Using the objective function~\eqref{eq:Prob_eachDU_est_obj} and the equality constraint in~\eqref{eq:Prob_eachDU_est_form}, we define the Lagrangian formulation as
\begin{align}\label{eq:FwithEqualityCon_DU_est}
	&f_1\left( {\bf W}_r, {\bm \xi}_r, {\bm \nu}_r \right) =
	\sum_{u \in \mathcal{E}_r} \delta_{u} \log\left( 1 + \xi_{ru} \right)
	\nonumber \\
	& \quad \quad \quad
	- \sum_{u \in \mathcal{E}_r} \nu_{ru} \left( \xi_{ru}
	- \frac{ s_{ru}
		{\bf w}_{ru}^H {\bf \hat{h}}_{ru} {\bf \hat{h}}_{ru}^H {\bf w}_{ru} 
	}
	{ 
		A_{ru}\left({\bf s}_r, {\bf W}_{r} \right)
	}
	\right)
	,
\end{align}
where ${\bm \xi}_r=[\{\xi_{ru}: u \in \mathcal{E}_r\}]^T \in \mathbb{R}^{|\mathcal{E}_r| \times 1}$. To satisfy the first-order optimality condition of $\xi_{ru}$, we set the derivative of~\eqref{eq:FwithEqualityCon_DU_est} with respect to $\xi_{ru}$ to zero, thereby obtaining an expression for $\nu_{ru}$ that satisfies this condition. Then, we substitute this expression for $\nu_{ru}$ back in~\eqref{eq:FwithEqualityCon_DU_est} to obtain 
\begin{align}\label{eq:Function1_DU_est}
	&f_1\left( {\bf s}_r, {\bf W}_r, {\bm \xi}_r \right) =
		\sum_{u \in \mathcal{E}_r}
		\delta_{u} \left( \log\left( 1 + \xi_{ru} \right) - \xi_{ru} \right)
		\nonumber \\
		&\ 
		+ \sum_{u \in \mathcal{E}_r}
		\delta_{u} \left( \frac{ \left( 1 + \xi_{ru} \right) s_{ru}
			{\bf w}_{ru}^H {\bf \hat{h}}_{ru} {\bf \hat{h}}_{ru}^H {\bf w}_{ru} 
		}
		{ 
			s_{ru}
			{\bf w}_{ru}^H {\bf \hat{h}}_{ru} {\bf \hat{h}}_{ru}^H {\bf w}_{ru}
			+ A_{ru}\left({\bf s}_r, {\bf W}_{r} \right)
		} \right)
	.
\end{align}
The importance of this procedure is that the variables $\left( {\bf s}_r, {\bf W}_r \right) $ are now found outside the logarithmic function, while $\xi_{ru}$ acts as an auxiliary variable or proxy to emulate their effect.
When the variables $\left( {\bf s}_r, {\bf W}_r \right) $ are fixed, we can set the derivative of~\eqref{eq:Function1_DU_est} with respect to $\xi_{ru}$ to zero, to obtain the optimal expression for the SLINR auxiliary variable $\xi_{ru}$, which, as expected, equals~\eqref{eq:Prob_eachDU_est_form}. The expression in~\eqref{eq:Function1_DU_est} can be shown to be equivalent to~\eqref{eq:FwithEqualityCon_DU_est} by substituting the optimal expression of $\xi_{ru}$ back into~\eqref{eq:Function1_DU_est}, thereby resulting in the same objective function in~\eqref{eq:Prob_eachDU_est}.

We use the fractional programming approach developed in~\cite{FR8310563} to write~\eqref{eq:Function1_DU_est} as
\begin{align}
	&f_2\left( {\bf s}_r, {\bf W}_r, {\bm \xi}_r, {\bm \zeta}_r \right) =
	\sum_{u \in \mathcal{E}_r}
	\delta_{u} \left( \log\left( 1 + \xi_{ru} \right) - \xi_{ru} \right)
	\nonumber \\
	&\ 
	+ \sum_{u \in \mathcal{E}_r}
	\bigg(
	2 \text{Re}\left\{
	\zeta_{ru}^{*}
	\sqrt{\delta_{u}\left( 1 + \xi_{ru}\right)}
	s_{ru}
	{\bf w}_{ru}^H {\bf \hat{h}}_{ru}
	\right\}
	\nonumber \\
	&\ 
	-
	|\zeta_{ru}|^2
	\left(
	s_{ru}
	{\bf w}_{ru}^H {\bf \hat{h}}_{ru} {\bf \hat{h}}_{ru}^H {\bf w}_{ru}
	+ A_{ru}\left({\bf s}_r, {\bf W}_{r} \right)
	\right)
	\bigg)
	.
\end{align}
where ${\bm \zeta}_r \in \mathbb{C}^{|\mathcal{E}_r| \times 1}$ is a new auxiliary variable vector introduced by fractional programming~\cite{FR8310563}. The formula in~\eqref{eq:Linearized_R} becomes the objective function of the newly formulated problem in~\eqref{eq:Prob_eachDU_est_reform}.

\section{Problem reformulation for CU-distributed system}
\label{eq:CUsystem_Form1}
Using the objective function~\eqref{eq:eq:optProbCU_v2_obj} and the equality constraint in~\eqref{eq:eq:optProbCU_v2_xi}, we can write the following Lagrangian formulation.
\begin{align}\label{eq:FwithEqualityCon_v2}
	&f_1\left( \mathcal{W}_q, \bar{\bm \xi}_{q}, {\bm \nu}_q \right) =
	\sum_{u \in \bar{\mathcal{U}}_{q} } \delta_{u} \log\left( 1 + \bar{\xi}_{qu} \right)
	\nonumber \\
	& \quad \quad \quad
	- \sum_{u \in \bar{\mathcal{U}}_{q} } \nu_{qu} \left( \bar{\xi}_{qu} -
	\frac{
		{\bf \bar{w}}_{qu}^H {\bf \bar{h}}_{qu,u} {\bf \bar{h}}_{qu,u}^H {\bf \bar{w}}_{qu} 
	}
	{ 
		C_{qu}\left(\mathcal{W}_{q} \right)
	} \right) .
\end{align}
To satisfy the first-order optimality condition of $\bar{\xi}_{qu}$, we differentiate~\eqref{eq:FwithEqualityCon_v2} with respect to $\bar{\xi}_{qu}$ and set to zero to obtain the corresponding value for $\nu_{qu}$. Substituting $\nu_{qu}$ back in~\eqref{eq:FwithEqualityCon_v2} we obtain 
\begin{align}\label{eq:Function1_v2}
	&f_1\left( \mathcal{W}_q, \bar{\bm \xi}_{q}\right) =
	\sum_{u \in \bar{\mathcal{U}}_{q} }
	\delta_{u} \left( \log\left( 1 + \bar{\xi}_{qu} \right) - \bar{\xi}_{qu} \right)
	\nonumber \\
	& \quad \quad
	+ \sum_{u \in \bar{\mathcal{U}}_{q} }
	\delta_{u} \left( \frac{ \left( 1 + \bar{\xi}_{qu} \right) 
		{\bf \bar{w}}_{qu}^H {\bf \bar{h}}_{qu,u} {\bf \bar{h}}_{qu,u}^H {\bf \bar{w}}_{qu} 
	}
	{ 
		{\bf \bar{w}}_{qu}^H {\bf \bar{h}}_{qu,u} {\bf \bar{h}}_{qu,u}^H {\bf \bar{w}}_{qu}
		+ C_{qu}\left(\mathcal{W}_{q} \right) 
	} \right) .
\end{align}
When $\mathcal{W}_q$ is fixed, the first optimality condition for $\bar{\xi}_{qu}$ yields the optimal value of $\bar{\xi}_{qu}$ as
\begin{align}\label{eq:optimalL_v2}
	\bar{\xi}_{qu}
	=
	\frac{ 
		{\bf \bar{w}}_{qu}^H {\bf \bar{h}}_{qu,u} {\bf \bar{h}}_{qu,u}^H {\bf \bar{w}}_{qu}
	}
	{ 
		C_{qu}\left(\mathcal{W}_{q} \right)
	} .
\end{align}
Equation~\eqref{eq:Function1_v2} can be shown to be equivalent to~\eqref{eq:FwithEqualityCon_v2} as discussed in the previous section for a DU-distributed system. Again, using fractional programming~\cite{FR8310563}, we write~\eqref{eq:Function1_v2} as
\begin{align}\label{eq:CUNewObj_v2_appendix}
	& f_2\left( \mathcal{W}_q, \bar{\bm \xi}_{q}, \bar{\bm \zeta}_q \right) =
	\sum_{u \in \bar{\mathcal{U}}_{q} }
	\delta_{u} \left( \log\left( 1 + \bar{\xi}_{qu} \right) - \bar{\xi}_{qu} \right)
	\nonumber \\
	& \ 
	+
	\sum_{u \in \bar{\mathcal{U}}_{q} }
	\Bigg(
	2 \text{Re}\left\{
	\bar{\zeta}_{qu}^{*}
	\sqrt{\delta_{u}\left( 1 + \bar{\xi}_{qu}\right)}
	{\bf \bar{w}}_{qu}^H {\bf \bar{h}}_{qu,u}
	\right\}
	\nonumber \\
	& \quad
	-
	|\bar{\zeta}_{qu}|^2
	\left(
	{\bf \bar{w}}_{qu}^H {\bf \bar{h}}_{qu,u} {\bf \bar{h}}_{qu,u}^H {\bf \bar{w}}_{qu}
	+ C_{qu}\left(\mathcal{W}_{q} \right)
	\right)
	\Bigg) ,
\end{align}
where $\bar{\bm \zeta}_{q} = [\bar{\zeta}_{qu_1}\ \dots\  \bar{\zeta}_{qu_{|\mathcal{U}_q|}}]^T$ are introduced auxiliary variables. The formula in~\eqref{eq:CUNewObj_v2_appendix} becomes the new objective function of our reformulated problem in~\eqref{eq:Prob_eachDU_est_reform}.

\footnotesize
\bibliography{RA_SLNR_References}
\bibliographystyle{ieeetr}

\end{document}